\documentclass[aps, prd, preprint, amsfonts,
  amssymb, amsmath, showpacs, a4paper, nofootinbib, superscriptaddress]{revtex4-1}

\usepackage{braket,slashed,bm}
\usepackage[dvipsnames]{xcolor}
\usepackage{graphicx}
\usepackage{soul}
\usepackage{mathtools}

 \usepackage{cellspace}
\usepackage[caption=false]{subfig}

\usepackage{array}
\usepackage{multirow}

\usepackage{natbib}
\usepackage{hyperref}
\newcommand{\phienics}{$\varphi$\texttt{enics}}
\newcommand{\fenics}{\texttt{FEniCS}}

\newcommand{\github}[1]{\url{https://github.com/benjthrussell}}

\begin{document}

\title{Fifth-Force Screening around Extremely Compact Sources}

\author{Clare Burrage}
\email{clare.burrage@nottingham.ac.uk}
\affiliation{School of Physics and Astronomy, University of Nottingham,\\
University Park, Nottingham NG7 2RD, United Kingdom}

\author{Benjamin Elder}
\email{bcelder@hawaii.edu}
\affiliation{School of Physics and Astronomy, University of Nottingham,\\
University Park, Nottingham NG7 2RD, United Kingdom}
\affiliation{Department of Physics and Astronomy,
University of Hawai'i,\\ Watanabe Hall,
2505 Correa Road, Honolulu, HI, 96822, USA}

\author{Peter Millington}
\email{p.millington@nottingham.ac.uk}
\affiliation{School of Physics and Astronomy, University of Nottingham,\\
University Park, Nottingham NG7 2RD, United Kingdom}

\author{Daniela Saadeh}
\email{daniela.saadeh@port.ac.uk}
\affiliation{School of Physics and Astronomy, University of Nottingham,\\
University Park, Nottingham NG7 2RD, United Kingdom}
\affiliation{Institute of Cosmology and Gravitation, University of Portsmouth,\\ Dennis Sciama Building, Portsmouth, PO1 3FX, United Kingdom}

\author{Ben Thrussell}
\email{benjamin.thrussell@nottingham.ac.uk}
\affiliation{School of Physics and Astronomy, University of Nottingham,\\
University Park, Nottingham NG7 2RD, United Kingdom}


\begin{abstract}
 Many non-linear scalar field theories possess a screening mechanism that can suppress any associated fifth force in dense environments.  As a result, these theories can evade local experimental tests of new forces.  Chameleon-like screening, which occurs because of non-linearities in the scalar potential or the coupling to matter, is well understood around extended objects.  However, many experimental tests of these theories involve objects with spatial extent much smaller than the scalar field's Compton wavelength, and which could therefore be considered point-like.  In this work, we determine how the fifth forces are screened in the limit that the source objects become extremely compact.   
\end{abstract}

\maketitle

\tableofcontents


\section{Introduction}

Non-linear scalar field theories are popular candidates  to help explain many cosmological mysteries, including  dark matter~\cite{Bertone:2004pz, Hui:2016ltb},  dark energy~\cite{Joyce:2014kja} and inflation~\cite{Akrami:2018odb}.  They also appear in many  modified theories of gravity, which often introduce new scalar degrees of freedom \cite{Joyce:2014kja} --- so called scalar-tensor theories.  Of course, the Standard Model Higgs sector  \cite{Englert:1964et, Higgs:1964pj, Guralnik:1964eu} is also described by a non-linear scalar field theory, although the energy scales involved are significantly higher than those of interest to late-universe cosmology.

In the absence of a symmetry principle~\cite{Shaposhnikov:2008xb, Brax:2014baa, Ferreira:2016kxi, Burrage:2018dvt}, it is difficult to forbid a new light scalar from coupling to matter, and, as a result, we expect it to mediate a long-range fifth force.   A canonical scalar field with a linear equation of motion has its mass and matter coupling fixed, and thus the existence of such forces is difficult to reconcile with terrestrial and Solar System tests of gravity~\cite{Adelberger:2009zz}.  However, the phenomenology of non-linear scalar field theories is much broader and allows for more interesting behaviour.  If the scalar theory is non-linear then the behaviour of fluctuations of the field, which mediate the fifth force, will depend on the properties of the background scalar evolution. In this way, the properties of the force, such as its coupling strength and range, can vary with the environment.  Of particular interest are theories in which the fifth force is dynamically suppressed in and around dense and compact sources.  This phenomenon is known as \emph{screening}, and such screening mechanisms can explain why the effects of new forces might evade laboratory experiments or in precision Solar System tests of gravity.

In this work, we focus on theories that screen through non-linearities in the field equations for the scalar $\phi$. The non-linearities may be terms in the scalar potential, in the way the field couples to matter or both. They should be contrasted with the non-canonical kinetic terms that effect Vainshtein screening~\cite{Vainshtein:1972sx}. The two key benchmark models relevant to this paper are the chameleon model~\cite{Khoury:2003aq, Khoury:2003rn, Mota:2006fz}, which relies on non-linearities in the potential to make fluctuations more massive in dense environments, and the symmetron model~\cite{Hinterbichler:2010es, Hinterbichler:2011ca} (for related earlier work, see Refs.~\cite{Dehnen:1992rr, Gessner:1992flm, Damour:1994zq, Pietroni:2005pv, Olive:2007aj}), which instead relies on a non-linear potential and a quadratic coupling to matter to vary both the mass of the field and the strength of the coupling to matter depending on the local energy density.

For screening models of this type, we know that, in any given environment, compact objects will either be screened, meaning the fifth force that they source or experience will be weaker than gravity, or unscreened.  This can be understood in terms of the scalar ``charges'' of these objects.  If the object is screened then the charge sourcing the scalar field around it will be smaller than the object's gravitational mass due to the so-called ``thin-shell effect''~\cite{Khoury:2003rn}.  If the object is unscreened then the scalar charge will correspond to the gravitational mass. 

The condition that determines whether or not an object is screened is typically phrased in terms of the object's mass (or density) and size.  However, in many situations, there is a large hierarchy between the (small) spatial extent of the source object and the (large) Compton wavelength of the scalar field.  We would like to be able to treat these sources as effectively point-like, but, to understand their screening properties, it is currently necessary to know the details of their structure. A deeper understanding of screening for compact and point-like sources is particularly relevant to laboratory experiments that use small test particles, such as atoms \cite{Hamilton:2015zga, Elder:2016yxm,Sabulsky:2018jma}, neutrons \cite{Lemmel:2015kwa,Cronenberg:2018qxf}  or micro-spheres \cite{Rider:2016xaq}, where the scalar Compton wavelength may be of order the size of the vacuum chamber; and to astrophysical environments, where we need to know the screening properties of stars and planets, but the scalar Compton wavelength may be of order galactic scales~\cite{Burrage:2017qrf}. Herein, we have differentiated between \emph{point-like} and \emph{compact} sources. While the former refers only to the spatial extent of the source, we consider a compact source to be one that is both small, compared to the Compton wavelength of the field, and high density.

A comprehensive treatment of the screening of point-like objects also represents an important first step towards a robust understanding of when distributions of discrete sources may be ``averaged over'' and treated as a single extended source with some continuous density. Such an understanding is also important for knowing how to include the effects of screening when computing interparticle forces in N-body simulations, e.g., of structure formation in the universe \cite{Llinares:2013jza, Winther:2014cia, Winther:2015wla, Bose:2016wms, Llinares:2018maz, Arnold:2019zup}. In this work, we consider the screening of fifth forces around compact, point-like objects in flat space, paving the way to further study in curved spacetimes, such as around black holes~\cite{Davis:2014tea, Davis:2016avf, Wong:2019yoc, Lagos:2020mzy}. 

Theories with screening have been studied in a wide variety of experimental environments~\cite{Joyce:2014kja, Upadhye:2012qu, Upadhye:2012rc, Hamilton:2015zga, Elder:2016yxm, Burrage:2017qrf, Sabulsky:2018jma, Rider:2016xaq,Lemmel:2015kwa, Cronenberg:2018qxf, Pernot-Borras:2019gqs, Stadnik:2020bfk} but, precisely because of the non-linearities, it is not always obvious how the behaviours of the field and the associated fifth force extrapolate to new regimes. Exact analytic solutions can often be obtained for systems that vary only in one spatial dimension, having applications to experiments that take place between parallel plates, such as torsion-pendulum, Casimir, and neutron experiments~\cite{Upadhye:2012rc, Brax:2013cfa, Brax:2017hna, Llinares:2018mzl, Pitschmann:2020ejb}, as well as the nested cylinders of the MICROSCOPE mission~\cite{Pernot-Borras:2020jev}. The solution for a point source in one spatial dimension is also known  analytically for particular choices of potential and coupling~\cite{Burrage:2018xta}.  Beyond these examples, however, our knowledge is more limited.  Therefore, in this work, we perform a detailed analysis of the impact of the dimensionality of the system on the problem of taking the point-like limit of an object in models with screening.

In this article, we aim to solve for the static configurations of the scalar field around spherical and cylindrical matter sources in the limit where the radius of the source tends to zero. We will see that it is important to differentiate between the point-like limit where the spatial extent of a source is sent to zero, while holding its mass fixed, and the compact limit where the size of the source is shrunk and the mass of the source is increased. Non-trivial static vacuum solutions of non-linear classical field theories are well known, and they are often referred to as (topological) defects.  These solutions may be stable if they're supported by some non-trivial topology of the vacuum manifold in field space~\cite{Vilenkin:2000jqa}, and their stability properties can be characterised using Derrick's theorem~\cite{Derrick:1964ww}. The solutions that we consider in this work differ in that they are supported by matter distributions.  We present a new extension of Derrick's theorem, which, for the first time, characterises the stability properties of field profiles supported by sources. Moreover, in three spatial dimensions, it allows us to infer the behaviour of the scalar field profiles in the point-like limit. While we might expect the point-like limit to be divergent, carefully studying the limiting procedure allows us to understand how these divergences are regulated\footnote{A similar approach is taken in theories of infinitesimally thin branes in higher-dimensional spacetimes~\cite{Goldberger:2001tn, deRham:2007mcp}.} and to determine how the fifth forces due to compact, point-like sources are screened.

The key results of this article are:
\begin{itemize}

    \item  An extension of Derrick's theorem to scalar profiles supported by external sources. 
    
    \item Piecewise analytic approximations for finding the field profiles around point-like sources.
    
    \item A numerical treatment of the scalar field profiles around point-like and compact sources.
    
    \item A comparison of the screening factors  between the full numerically calculated field profiles and the approximate solutions that are currently used in the literature to quantify the amount by which the fifth force is screened.

\end{itemize}
A summary of the behaviour of screening in this limit for the two prototypical theories that we study, and for spherically and cylindrically symmetric sources, can be found in Table~\ref{table:scalings}.

This paper is organised as follows. Section~\ref{section:models} provides an overview of the particular models that we will examine in later sections, as well as an introduction to the screening factors that have previously appeared in the literature. Section~\ref{section:Derrick} describes how Derrick's theorem is modified in the presence of classical sources and discusses its implications for the behaviour of the scalar field profiles as the source becomes increasingly compact. In Sec.~\ref{sec:fieldprofiles}, we turn our attention to the computation of the field profiles.  Section~\ref{section:piecewisepotentials} focuses on improved piecewise approximations to the potentials and field profiles. Section~\ref{section:numerics} subsequently describes the numerical procedure that we employ to solve for the field profile without making any approximations to the form of its potential. In Sec.~\ref{sec:implications}, we describe the implications of our numerical results for screening and estimates of screening around extremely compact sources. In Sec.~\ref{section:quantum}, we comment on the reliability of our classical calculations, establish the regimes in which we trust the classical predictions and identify when quantum corrections may become large. Finally, we provide our conclusions in Sec.~\ref{sec:conclusions}.  Extra details relating to the numerical procedures and an explicit example of a construction of a piecewise field profile are included in the appendices.

Throughout this work, we use the mostly plus metric signature convention $(-,+,+,+)$.  A summary of our notation is provided in Table~\ref{tab:notation}.

\begingroup
\renewcommand*{\arraystretch}{1}
\renewcommand*{\baselinestretch}{1}
\begin{table}[t]
    \centering
    \begin{tabular}{|@{\hspace{0.5em}}c@{\hspace{0.5em}}| @{\hspace{0.5em}}p{12cm}@{\hspace{0.5em}}|}
    \hline
        $M$ & mass scale determining the strength of the matter coupling \\
        $\mu$ & mass parameter in the potential of the screening scalar\\
        $\alpha$ & quartic self-coupling of the screening scalar\\
        $\rho$ & energy density of the matter source \\
        $\lambda_s$ & screening factor \\
        $\bar{\phi}(\rho)$ & position of the density-dependent minimum of the effective potential\\
        $\bar{\phi}_{\infty}\equiv\bar{\phi}(0)$ & position of the minimum of the effective potential for $\rho =0$\\
        $r_s$ & source radius\\
        $m_s$ & source mass\\
        $\rho_{\rm in}$ & constant density of the source \\
        $\rho_{\rm out}$ & constant ambient density external to the source\\
        $\bar{\phi}_{\rm in(out)}\equiv\bar{\phi}(\rho_{\rm in(out)})$ & position of the minimum of the effective potential for $\rho=\rho_{\rm in(out)}$\\
        $m_{\rm in(out)}$ & effective mass of the scalar fluctuations inside(outside) the source\\
        $x\equiv \mu r$ & dimensionless radius\\
        $x_s\equiv \mu r_s$ & dimensionless source radius \\
        $r_{\rm shell}$ & thin-shell radius\\
        $d$ & number of spatial dimensions\\
        $\tilde{\rho}\equiv\rho/\rho_c$ & normalised density:~$\rho_c=\mu^2M^2$ for the quadratically coupled model and $\rho_c=\mu^3M/\sqrt{\alpha}$ for the linearly coupled model\\
        $\tilde{m}_s$ & normalised source mass, $\tilde{m}_s=m_s\mu^{d-2}\bar{\phi}_{\infty}^{n-2}/M^n$:~for the quadratically coupled model ($n=2)$, $\tilde{m}_s=m_s\mu^{d-2}/M^2$; for the linearly coupled model ($n=1$), $\tilde{m}_s=m_s\sqrt{\alpha}\mu^{d-3}/M$ \\
        $\nu$ & effective mass of scalar fluctuations in units of $\mu$\\
        $\lambda_{\rm approx}$ & screening factor as obtained from piecewise quadratic approximations to the effective potential\\
        $\lambda_{\rm full}$ & screening factor as obtained from the full numerical solutions\\
        \hline
    \end{tabular}
    \caption{Summary of key notation and their definitions (in order of appearance).}
    \label{tab:notation}
\end{table}
\endgroup


\section{Screening models}
\label{section:models}

In this section, we introduce the scalar-tensor theories that will be the focus of this article.  The fifth forces that the additional scalar fields mediate are subject to screening mechanisms, which arise due to non-linear terms in their ``effective potentials''.\footnote{The term ``effective potential'' used here should not be confused with the effective potential that includes loop corrections to the tree-level potential.}

The models that we consider have an action of the form
\begin{equation}
\label{eq:STtheoriesgeneral}
S=\int d^4 x\sqrt{-g}\left[\frac{R}{16\pi G}-\frac{1}{2}\nabla_\mu\phi\nabla^\mu\phi-V(\phi)\right]+S_{m}[\tilde{g}_{\mu\nu},\{\psi\}],
\end{equation}
which describes a canonically normalised scalar field $\phi$ that is subject to a potential $V(\phi)$ and conformally (Weyl) coupled to matter, which we denote by the set of fields $\{\psi\}$. Herein, $R$ is the Ricci scalar and $G$ is Newton's gravitational constant. Matter fields move on geodesics of the Jordan-frame metric
\begin{equation}
\label{eq:Weylgen}
\tilde{g}_{\mu\nu}=A^2(\phi)g_{\mu\nu},
\end{equation}
coupling to the field $\phi$ through the coupling function $A(\phi)$.

Since we will focus on solutions in Minkowski spacetime, we have chosen to work in the so-called Einstein frame, wherein the gravitational part of the action has a canonical Einstein-Hilbert form, and the canonically normalised scalar field $\phi$ is coupled directly to the trace of the matter energy-momentum tensor. Alternatively, we could have worked in the Jordan frame, wherein the scalar field is coupled directly to the Ricci scalar, and the matter coupling is not explicit. The Jordan and Einstein frames are related by the Weyl rescaling of the metric that appears in Eq.~\eqref{eq:Weylgen}.

The coupling function $A(\phi)$ is dimensionless. Hence, in $3+1$ dimensions, wherein the scalar field has mass dimension one, this coupling function must also depend on one or more mass scales, which together determine the strengths of the couplings to matter (or, equivalently, the strength of the non-minimal coupling to the gravity sector). Making the minimal assumption of a single mass scale $M$ and restricting ourselves to the regime $\phi/M\ll 1$, we can expand the coupling function as a power series in $\phi/M$. In general, this series will yield non-renormalisable operators, and we therefore have an effective field theory with a cut-off scale parametrically related to $M$. In what follows, we keep terms suppressed by up to and including two powers of the scale $M$.

Keeping only these terms, the coupling functions of the models that we study are
\begin{equation}
    A_{\rm lin.}(\phi)= 1 + \frac{\phi}{M}~,
    \label{eq:lin}
\end{equation}
and
\begin{equation}
    A_{\rm quad.}(\phi)= 1 + \frac{\phi^2}{2M^2}~,
    \label{eq:quad}
\end{equation}
leading to linear and quadratic couplings to matter, respectively. A linear coupling to matter arises in the chameleon model~\cite{Khoury:2003aq}, whereas the quadratic coupling is used in the symmetron model~\cite{Hinterbichler:2010es}. The omission of a linear term in $\phi/M$  in Eq.~\eqref{eq:quad} follows directly from requiring a $\mathbb{Z}_2$ symmetry of the coupling function, i.e., a symmetry under $\phi\to-\phi$.

Because of the key role that non-linear effects play in their behaviour, we consider theories that exhibit spontaneous symmetry breaking (SSB), akin to the symmetron model~\cite{Hinterbichler:2010es}. We take the potential for the scalar field to be of the form: 
\begin{equation}\label{eq:symmetronpotential}
V(\phi) = - \frac{\mu^2}{2}\phi^2+\frac{\alpha}{4}\phi^4 + \frac{1}{4}\,\frac{\mu^4}{\alpha}~,
\end{equation}
where the dimensionless self-coupling $\alpha>0$ and squared mass parameter $\mu^2>0$ are real. The constant $\mu^4/(4\alpha)$  ensures that the field has zero energy density in the Minkowski vacuum. While this constant is irrelevant from the point of view of the scalar-field equations of motion, it is relevant for the applicability of Derrick's thoerem, as described in Sec.~\ref{section:Derrick}.  Spontaneous symmetry breaking can also be realised in the context of symmetron screening~\cite{Burrage:2016xzz} via the Coleman-Weinberg mechanism~\cite{Coleman:1973jx}, and we expect the behaviour of this variant to be qualitatively similar.

It is possible to find analytic solutions for models with the potential in Eq.~\eqref{eq:symmetronpotential} in one spatial dimension only. Even a semi-analytic analysis of the behaviour of the solutions around maximally symmetric sources in higher dimensions is extremely challenging. Given the importance of non-linear effects, these models therefore provide an illustrative proving ground within which to develop and apply the techniques discussed in this work.

The coupling of the scalar field to matter means that its dynamics can be understood as being due to an ``effective potential''
\begin{equation}
    V_{\rm eff}(\phi)= V(\phi) - A(\phi) T_{~\mu}^{\mu}~,
\end{equation}
where $T_{~\mu}^{\mu}$ is the trace of the covariantly conserved energy-momentum tensor of matter (see, e.g., Ref.~\cite{Joyce:2014kja}).  In non-relativistic environments, matter can be treated as a perfect fluid, and we can write $T_{~\mu}^{\mu}=-\rho$, where $\rho$ is the energy density of matter.

The coupling of the scalar field to matter fields leads to two effects. First, matter fields act, through the trace of their energy-momentum tensor, as a source for the scalar field. Second, gradients in the scalar field give rise to a ``fifth force'', such that matter does not move on geodesics of the Einstein-frame metric. The scalar fifth force per unit mass has the form 
\begin{equation}
    \vec{F}_{\phi}= -\frac{1}{A(\phi)} \vec{\nabla}A(\phi)~,
    \label{eq:force}
\end{equation}
manifesting as a deviation from general relativity. As discussed in the introduction, the non-linearities of the scalar field can make the strength of this fifth force vary depending on the environment, and the effects of the fifth force can be suppressed or \emph{screened}.

The scalar field theories that we consider in this work have a non-vanishing mass $m_{\rm out}$ exterior to the source. As a result, the fifth force mediated by the scalar fluctuations is subject to a Yukawa suppression at radii larger than $1/m_{\rm out}$. However, the focus of the present work is the additional suppression of the fifth force due to non-linear effects in addition to the Yukawa suppression.

In order to quantify screening, it is common to introduce a ``screening factor'' $\lambda_s$, and a number of different definitions of this factor have appeared in the literature; see, e.g., Refs.~\cite{Khoury:2003rn,Mota:2006fz}. In phenomenological applications, this screening factor is often defined in terms of the ratio of the magnitudes of the scalar fifth force $F_{\phi}$ to the Newtonian gravitational force $F_{\rm N}$. However, we instead wish to use a screening factor in which the impact of non-linearities is not confounded with the Yukawa suppression described above. To this end, we define
\begin{equation}
    \lambda_s=\frac{F_{\phi}}{F_{\rm N}}.\frac{F_{\rm N}}{F_{\rm Yukawa}}=\frac{F_{\phi}}{F_{\rm Yukawa}}~,
    \label{eq:screeningfac}
\end{equation}
where $F_{\rm Yukawa}$ is the magnitude of the Yukawa force due to scalar fluctuations of the same (effective) mass but with a quadratic potential. Well inside the Compton wavelength of the scalar field, the screening factor in Eq.~\eqref{eq:screeningfac} is directly proportional to previous definitions in terms of $F_{\phi}/F_N$.

We will now review the standard argument for how this screening takes places around static, compact and spherically symmetric sources, focusing on the quadratically and linearly coupled models described above.  Here, ``linearly coupled'' and ``quadratically coupled'' refer to the coupling as it appears in the effective potential, not as it appears in the equation of motion.


\subsection{Quadratically coupled model}
\label{section:quadscreening}

Taking the matter degrees of freedom to be non-relativistic, the effective potential of the \emph{quadratically} coupled model, with the coupling function given in Eq.~\eqref{eq:quad}, takes the form
\begin{equation}
\label{eq:quadpotential}
V_{\rm eff}(\phi)=V(\phi)+\frac{1}{2}\frac{\rho}{M^2}\phi^2~.
\end{equation}
When the local density $\rho \geq \mu^2 M^2$, the effective potential only has one minimum $\bar{\phi}(\rho)$ at $\bar{\phi}(\rho) = 0$.  In regions of lower density with $\rho < \mu^2 M^2$, the effective potential  has two minima at $\bar{\phi}(\rho) = \pm \bar{\phi}_{\infty}\sqrt{1 -\rho/\mu^2M^2}$, where $\bar{\phi}_{\infty} = \mu/\sqrt{\alpha}$ is the field value at the minimum of the effective potential when $\rho =0$.  Since the classical fifth force given by Eq.~\eqref{eq:force} is proportional to the ambient field value, we have $\bar \phi(\rho) = 0$ in sufficiently dense regions and the fifth force vanishes.

In order to treat the screening behaviour analytically, we consider a compact, maximally symmetric source with finite radius $r_s$, uniform density $\rho_{\rm in}$ and total mass $m_s$, embedded in a diffuse background with uniform density $\rho_{\rm out}$ that is less than the critical value $\mu^2M^2$. In the literature, e.g., Ref.~\cite{Hinterbichler:2011ca}, the screening of the fifth force sourced by this object is determined by dividing space into two regions:~one inside and one outside the source object. In addition, we make the following assumptions:~(i) the field inside the source remains close to zero and (ii) the field outside the source remains close to the value that minimises the effective potential outside the source, which we denote by
\begin{equation}
\bar{\phi}_{\rm out} \equiv \bar{\phi}(\rho_{\rm out})=\bar{\phi}_{\infty}\sqrt{1 -\rho_{\rm out}/\mu^2M^2}~,
\end{equation}
where we have arbitrarily (but without loss of generality) selected the positive vacuum solution at spatial infinity. We then approximate the effective potential as quadratic about its minimum in each of the two regions.  The effective mass of the field inside and outside the source is given by
\begin{equation}
    m_{{\rm in}({\rm out})}^2= \left.\frac{{\rm d}^2V_{\rm eff}(\phi)}{{\rm  d}\phi^2}\right|_{\substack{\phi\,=\,\bar{\phi}_{{\rm in}({\rm out})}\\ \rho\,=\,\rho_{{\rm in}({\rm out})}}}~,
\end{equation}
where $\bar{\phi}_{\rm in}=\bar{\phi}(\rho_{\rm in})$. We impose the boundary conditions that the field be regular at the origin and tend to a constant at infinity, and that $\phi$ and ${\rm d} \phi/ {\rm d}r$ be continuous at the surface of the source.

After making the assumptions described above and further approximating the fifth force as $\vec{F}=-\phi \vec \nabla \phi/M^2 \approx -\bar \phi_\mathrm{out} \vec \nabla \phi/M^2$,\footnote{The full expression for $\lambda_s$ does not approximate the fifth force $-\phi\vec{\nabla} \phi/M^2$ as $-\bar{\phi}_{\rm out}\vec{\nabla}\phi/M^2$ and therefore depends on the distance from the source.} the external field profile around our object, which we centre at the origin, can be written as \cite{Burrage:2016rkv}
\begin{equation}\label{eq:3Dquadphiout}
\phi_{\rm out}(r) = \bar{\phi}_{\rm out} - \lambda_s \frac{m_s}{\bar{\phi}_{\rm out}}\frac{e^{-m_{\rm out}(r-r_s)}}{(1+m_{\rm out}r_s)4\pi r}~,
\end{equation}
where the screening factor $\lambda_s$ is given by
\begin{equation}\label{eq:3Dquadlam}
\lambda_s = \frac{4\pi\bar{\phi}_{\rm out}^2(1+m_{\rm out}r_s)}{m_s}\left(1 - \frac{\bar{\phi}_{\rm in}}{\bar{\phi}_{\rm out}}\right)\frac{m_{\rm in}r_s - \tanh(m_{\rm in}r_s)}{m_{\rm in} + m_{\rm out}\tanh(m_{\rm in}r_s)}~.
\end{equation}

The cylindrically symmetric (i.e., two-dimensional) case can be treated in a similar manner, with the exterior field falling off as the modified Bessel function of the second kind $K_0(m_{\rm{out}}r)$:
\begin{equation}
\phi_{\rm out}(r) = \bar{\phi}_{\rm out} - \lambda_s\frac{m_s}{\bar{\phi}_{\rm out}}\frac{1}{2\pi m_{\rm out}r_s}\frac{K_0(m_{\rm out}r)}{K_1(m_{\rm out}r_s)}~,
\end{equation}
with
\begin{equation}
\lambda_s = \frac{2\pi\bar{\phi}_{\rm out}^2}{m_s}\left(1 - \frac{\bar{\phi}_{\rm in}}{\bar{\phi}_{\rm out}}\right)\frac{m_{\rm in}m_{\rm out}r_s I_1(m_{\rm in}r_s)K_1(m_{\rm out}r_s)}{m_{\rm in} I_1(m_{\rm in}r_s)K_0(m_{\rm out}r_s) + m_{\rm out} I_0(m_{\rm in}r_s)K_1(m_{\rm out}r_s)}~,
\end{equation}
where we have again assumed that the fifth force can be approximated as $-\bar{\phi}_{\rm out}\vec{\nabla}\phi/M^2$.

Lastly, we review the one-dimensional case that was analysed in detail in Ref.~\cite{Burrage:2018xta}. For the quadratically coupled model, the equation of motion takes the form
\begin{equation}
    \varphi''(x)+\varphi(x)[1-\varphi^2(x)]=\begin{cases}\frac{\rho}{\mu^2M^2}\varphi(x)~,&\qquad x\leq x_s\\
    0~,&\qquad x>x_s\end{cases}~,
\end{equation}
where $\varphi=\phi/\bar{\phi}_{\infty}$, $x=\mu r$, $x_s=\mu r_s$ and primes indicate derivatives with respect to $x$. The exterior solution is
\begin{equation}
    \label{eq:1Dext}
    \varphi(|x|>x_s)=\tanh\left(\frac{1}{\sqrt{2}}(|x|-x_s)+{\rm arctanh}\,\varphi(x_s)\right)~,
\end{equation}
and the interior solution can be written in terms of the Jacobi elliptic function ${\rm nc}$ as
\begin{equation}
    \varphi(|x|\leq x_s)=\varphi(0)\,{\rm nc}\left(\gamma x,1-\frac{\varphi(0)^2}{2\gamma^2}\right)~,
\end{equation}
where
\begin{equation}
\gamma=\sqrt{\frac{\rho}{\mu^2M^2}-1+\varphi(0)^2}~.
\end{equation}
The values of the field $\varphi$ at the origin and at the surface of the source, $\varphi(0)$ and $\varphi(x_s)$ respectively, are obtained by matching the solutions and their first derivatives at $x_s$, which can be done only semi-analytically for a source of finite radius.

In the point-source limit, $x_s\to0$, we have $\rho(x)=m_s\delta(x)$, and the solution reduces to
\begin{equation}
\varphi(x)=\tanh\left(\frac{1}{\sqrt{2}}|x|+{\rm arctanh}\,\varphi(0)\right)~,
\end{equation}
with
\begin{equation}
    \varphi(0)=\frac{1}{2\sqrt{2}}\left(\sqrt{\frac{m_s^2}{\mu^4M^4}+8}-\frac{m_s}{\mu^2 M^2}\right)~.
\end{equation}
The screening factor for a point source is thus
\begin{equation}
    \lambda_s= \frac{4 \mu \bar{\phi}_{\infty}^2}{m_s}\frac{1-\phi(0)}{1+\phi(0)}~,
\end{equation}
with the limiting behaviour
\begin{equation}
    \lambda_s =\frac{\bar{\phi}_{\infty}^2}{\sqrt{2}M^2} \begin{cases}
    1~, & m_s \ll \mu M^2\\
    \frac{\mu M^2}{m_s}~, & m_s \gg \mu M^2
    \end{cases}~,
\end{equation}
and we see that the fifth forces sourced by more massive objects are screened.


\subsection{Linearly coupled model}
\label{section:linscreening}

For the case where the scalar coupling to matter is linear [cf.~Eq.~\eqref{eq:lin}], we again take the matter degrees of freedom to be non-relativistic so that the effective potential is
\begin{equation}\label{eq:linpotential}
V_{\rm eff}(\phi)=V(\phi)+\frac{\rho}{M}\phi~.
\end{equation}
Unlike the quadratically coupled model, there is no critical density or phase transition between screened and unscreened regimes.  In this case, the mass of fluctuations in the scalar field increases with the ambient matter density.  This makes the perturbations, and hence the force, sourced by regions deep inside a large dense object short-ranged~\cite{Mota:2006fz}~.

Before continuing, we remark further on our truncation of the expansion of $A(\phi)$ to leading order in $\phi/M\ll 1$.  At leading order, we have dropped the term $\rho \phi^2/(2M^2)$, since it is subleading compared to $\rho\phi/M$.  However, the power counting in $\phi/M\ll 1$ does not ensure that $\rho\phi^2/(2M^2)$ is subleading compared to the mass term $-\mu^2\phi^2/2$.  In fact, this will not be the case for much of the parameter space that we study later in this work.  We therefore need to justify keeping the mass term in the potential, while dropping the quadratic term from the expansion of the coupling function.  One possibility would be to consider a coupling function with two scales:~$M_1$, which controls the odd power series in $\phi$, and $M_2$, which controls the even power series in $\phi$.\footnote{For example, such a coupling function might be given by $A_{\rm lin.}(\phi)=\sinh(\phi/M_1)+\cosh(\phi/M_2)$.} We can then arrange for the quadratic term in the expansion of $A(\phi)$ to be subleading compared to the mass term in $V(\phi)$ without affecting the magnitude of the linear coupling. Similar fine-tuning arguments apply in respect of the quartic operator in $V(\phi)$, and the cubic and quartic operators arising from the expansion of $A(\phi)$.

Considering again a uniform-density, maximally symmetric source of mass $m_s$ and radius $r_s$, embedded in a low-density background,  the standard approach to computing the screening factor is to divide space into three regions:~one outside and two inside the source~\cite{Khoury:2003rn}. In the innermost region, the field is so massive that the field value is approximately constant.  The surface of this region defines a ``thin shell'' radius $r_{\rm shell}$ with $0\leq r_{\rm shell}\leq r_s$. In the region $r_{\rm shell}\leq r\leq r_s$, the effective potential is dominated by the matter coupling.

To leading order in $\phi/M$,  the field value that minimises the effective potential inside the source must satisfy
\begin{equation}
    \alpha \bar{\phi}_{\rm in}^3-\mu^2\bar{\phi}_{\rm in}+\frac{\rho_{\rm in}}{M}=0~. 
\end{equation}
and the mass of the scalar fluctuations about $\bar{\phi}_{\rm in}$ is
\begin{equation}
    m_{\rm in}^2 = -\mu^2 +3 \alpha \bar{\phi}_{\rm in}^2~.
\end{equation}
The exterior field profile around a homogeneous spherically-symmetric source of radius $r_s$ is~\cite{Burrage:2014oza}
\begin{equation}\label{eq:3Dlinphiout}
\phi_{\rm out}(r) = \bar{\phi}_{\rm out} - \lambda_s\frac{m_s}{M}\frac{e^{-m_{\rm out}(r-r_s)}}{(1+m_{\rm out}r_s)4\pi r}~.
\end{equation}
The resulting screening factor is given by 
\begin{equation}\label{eq:3Dlinlam}
\lambda_s = 1 - \left(\frac{r_{\rm shell}}{r_s}\right)^3~
\end{equation}
and, taking $m_{\rm out} r \ll 1$, the thin-shell radius is
\begin{equation}\label{eq:3Drs}
r_{\rm shell} = r_s\sqrt{1-\frac{2M(\bar{\phi}_{\rm out}-\bar{\phi}_{\rm in})}{\rho_{\rm in} r_s^2}}~.
\end{equation}
Only the matter within the shell near the surface of the object sources perturbations in the field. This form of the solution holds for all models with a linear coupling to matter, as different self-interaction potentials $V(\phi)$ merely lead to different expressions for $\bar{\phi}_{\rm out}$.

When $r_{\rm shell} = 0$ and $\lambda_s = 1$, which occurs for $2M(\bar{\phi}_{\rm out}-\bar{\phi}_{\rm in}) \ge \rho_{\rm in} r_s^2$, the field cannot reach its minimum anywhere and the object is entirely unscreened. When $r_{\rm shell} \to r_s$ and $\lambda_s \to 0$, occurring when $\rho_{\rm in} r_s^2 \gg 2M(\bar{\phi}_{\rm out}-\bar{\phi}_{\rm in})$, the field quickly drops to its minimum near the surface of the source and the object is maximally screened.

The cylindrically symmetric or two-dimensional case is similar, with \cite{Burrage:2014oza}
\begin{equation}\label{eq:2Dlinphi}
\phi_{\rm out}(r) = \bar{\phi}_{\rm out} - \lambda_s\frac{m_s}{M}\frac{1}{2\pi m_{\rm out} r_s}\frac{K_0(m_{\rm out} r)}{K_1(m_{\rm out} r_s)}~,\qquad\lambda_s = 1 - \left(\frac{r_{\rm shell}}{r_s}\right)^2~,
\end{equation}
and the thin-shell radius $r_{\rm shell}$ is given by
\begin{equation}\label{2drs}
\left(\frac{r_{\rm shell}}{r_s}\right)^2 = \frac{-(X + Y)}{W_{-1}(-e^{-X}(X+Y))}~,
\end{equation}
where
\begin{equation}
X = 1 + \frac{2}{m_{\rm out}r_s}\frac{K_0(m_{\rm out}r_s)}{K_1(m_{\rm out}r_s)}~,\qquad Y = \frac{4M(\bar{\phi}_{\rm in} - \bar{\phi}_{\rm out})}{\rho r_s^2}
\end{equation}
and $W_k$ denotes the $k^{\rm th}$ real branch of the Lambert $W$ function. The argument of the $W$ function becomes non-negative when $X < -Y$, and Eq.~\eqref{2drs} no longer gives an $r_{\rm shell}$ between 0 and $r_s$. As the argument approaches zero, the $W$ function tends to $-\infty$, $r_{\rm shell} \to 0$, and the source is unscreened.

In the one-dimensional case, the equation of motion takes the form
\begin{equation}
    \varphi''(x)+\varphi(x)[1-\varphi^2(x)]=\begin{cases}\frac{\rho}{\mu^2 M\bar{\phi}_{\infty}}~,&\qquad x\leq x_s\\
    0~,&\qquad x>x_s\end{cases}~.
\end{equation}
The interior solution is not of a simple form, nor is it particularly illuminating, and we therefore omit it here. On the other hand, the exterior solution is the same as in Eq.~\eqref{eq:1Dext}, so, in the point-source limit, we have~\cite{Burrage:2018xta}
\begin{equation}
    \label{eq:oneDvarphi}
    \varphi(0)=\sqrt{1-\frac{m_s}{\sqrt{2}\mu M \bar{\phi}_{\infty}}}~.
\end{equation}
We encounter a branch point in the square root at $m_s=\sqrt{2}\mu M \bar{\phi}_{\infty}$, at which the solution becomes imaginary, signifying a breakdown phenomenon~\cite{Burrage:2018xta}. As we will see in Sec.~\ref{sec:fieldprofiles}, this behaviour is not observed in the linearly coupled model in spatial dimensions $d>1$~.

The screening factor of a point source is thus
\begin{equation}
    \lambda_s= \frac{2 \sqrt{2} \mu M \bar{\phi}_{\infty}}{m_s}\frac{1-\phi(0)}{1+\phi(0)}\;,
\end{equation}
with the limiting behaviour
\begin{equation}
    \lambda_s \approx 1\qquad \text{when}\qquad m_s \ll \mu M \bar{\phi}_{\infty}\;.
\end{equation}


\section{Derrick's theorem with classical sources}
\label{section:Derrick}

Derrick's theorem \cite{Derrick:1964ww} provides a set of conditions for the existence of stable and stationary localised solutions to non-linear field equations. It is usually applied in vacuum, that is, in the absence of sources. In this section, we generalise the derivation of Derrick's theorem to incorporate the presence of a classical hyperspherically, i.e., $O(d)$ symmetric uniform source, with density described by a top-hat distribution.  The usual stability arguments then translate into information about how the theories introduced in the previous section behave around point-like sources and allow us to test the accuracy of our numerical solutions in Sec.~\ref{section:numerics}. For comparison, the standard derivation of Derrick's theorem is recovered straightforwardly when the density $\rho$ of the classical source is taken to zero.

We are interested in  stationary localised solutions, which  must correspond to a minimum of the energy.  For the quadratically and linearly coupled models that we study here, the energy is given in $d$ spatial dimensions by
\begin{equation}
E[\phi]\ =\ \int{\rm d}^d\mathbf{x}\;\bigg[\frac{1}{2}\,\big(\bm{\nabla}\phi(\mathbf{x})\big)^2\:+\:V(\phi(\mathbf{x})) + \frac{1}{n!}\;\frac{\rho(\mathbf{x})}{M^n}\,\phi^{n}\bigg]~,
\end{equation}
where $n=1,2$. This energy is finite as long as we shift the potential so that the field has zero energy density in vacuum, see Eq.~\eqref{eq:symmetronpotential}. We can write the contributions from the kinetic, potential and matter-coupling parts respectively as
\begin{subequations}
\begin{align}
E_K[\phi]\ &\equiv\  \frac{1}{2}\int{\rm d}^d\mathbf{x}\;\big(\bm{\nabla}\phi(\mathbf{x})\big)^2~,\\
E_V[\phi]\ &\equiv\  \frac{1}{2}\int{\rm d}^d\mathbf{x}\;V(\phi(\mathbf{x}))~,\\
E_{\rho}[\phi]\ &\equiv\ \frac{1}{n!}\int{\rm d}^d\mathbf{x}\;\frac{\rho(\mathbf{x})}{M^n}\,\phi^{n}~.
\end{align}
\end{subequations}
For a spherically symmetric, uniform source with $\rho(\mathbf{x}) \equiv \rho(r)=\rho_{\rm in}\theta(r_s-r)$, where $\rho_{\rm in}$ is a constant and $\theta$ is the unit step function, we can write
\begin{equation}
E_{\rho}[\phi]\ =\ \frac{1}{n!}\int{\rm d}\Omega_{d}\int_0^{r_s}{\rm d}r\;r^{d-1}\,\frac{\rho_{\rm in}}{M^n}\,\phi^{n}(r)~,
\end{equation}
where $\int{\rm d}\Omega_{d}$ is the integral over the solid angle
\begin{equation}
    \label{eq:solidangle}
    \Omega_d=2\pi^{d/2}/\Gamma(d/2)
\end{equation}
subtended by the $d$-dimensional sphere.

In order to determine whether any given solution $\phi$ is stable (with respect to growth or collapse), we are interested in the first and second variations of the energy with respect to the transformation $\phi(\mathbf{x})\to \phi'(\mathbf{x})=\phi(\omega\mathbf{x})$, requiring that $\delta E[\phi']/\delta \omega\big|_{\omega\,=\,1}=0$ and $\delta^2 E[\phi']/\delta \omega^2\big|_{\omega\,=\,1}\geq 0$. These conditions are respectively
\begin{subequations}
\begin{gather}
\label{eq:Derrick1}
\frac{1}{n!}\,\Omega_d\,r_s^d\,\frac{\rho_{\rm in}}{M^n}\,\phi^n(r_s)\ =\ (d-2)E_K[\phi]\:+\:d\big(E_V[\phi]+E_{\rho}[\phi]\big)~,\\
\label{eq:Derrick2}
\frac{1}{(n-1)!}\,\Omega_d\,r_s^{1+d}\,\frac{\rho_{\rm in}}{M^n}\,\phi^{n-1}(r_s)\partial_{r_s}\phi(r_s)\ \geq\ 2(d-2)E_K[\phi]~.
\end{gather}
\end{subequations}
Writing $m_s = \rho_{\rm in}\Omega_d r_s^d/d$ (the source mass), we have
\begin{subequations}
\label{eq:Derricksourcefinal}
\begin{gather}
\label{eq:Derricksource1}
\frac{d}{n!}\,\frac{m_s}{M^n}\,\phi^n(r_s)\ =\ (d-2)E_K[\phi]\:+\:d\big(E_V[\phi]+E_{\rho}[\phi]\big)~,\\
\label{eq:Derricksource2}
\frac{d}{(n-1)!}\,r_s\,\frac{m_s}{M^n}\,\phi^{n-1}(r_s)\partial_{r_s}\phi(r_s)\ \geq\ 2(d-2)E_K[\phi]~.
\end{gather}
\end{subequations}

We now suppose that we want the field profile to be generated by a point source of finite mass, with $r_s \to 0$. As long as $\phi$ and its spatial derivatives remain finite, this removes the left-hand side of Eq.~\eqref{eq:Derricksource2}, giving $(d-2)E_K[\phi] \le 0$. Since $E_K[\phi] > 0$ for theories with a canonical kinetic term and spatially varying profiles, this is satisfied in $d = 1$ and $2$, but not in $d > 2$. Therefore, for any model with a canonical kinetic term, the derivative of any stable and stationary field has to diverge at $r_s$ in $d > 2$ as $r_s \to 0$.

Whether the field itself also diverges cannot be derived from the arguments above,  but we might expect it to depend on the form of the potential and the matter coupling, such that the field will tend towards the minimum of its effective potential. This may or may not diverge as $\rho \to \infty$, depending on the choice of $V(\phi)$ and the value of $n$. With the choice of bare potential in Eq.~\eqref{eq:symmetronpotential}, an even value of $n$ brings the minimum of $V_{\rm eff}(\phi)$ to zero as $\rho \to \infty$, while an odd value causes it to diverge, as we will see later in Sec.~\ref{section:numerics} (see Fig.~\ref{fig:numericprofiles}).

In the remainder of this work, we take $d$ to be the effective dimensionality of the source. The case $d=3$ corresponds to a spherically symmetric source in three spatial dimensions; the case $d=2$ instead corresponds to a cylindrically symmetric source in three spatial dimensions, i.e., a cylinder with infinite extent along its symmetry axis. For the latter case, we assume that the solution $\phi(r,z)$ in three spatial dimensions has a separable solution, $\phi(r,z)=\phi(r)Z(z)$ (with some abuse of notation), such that we can integrate out $Z(z)$ to obtain an effective two-dimensional action, absorbing the mass scales and separation constant into the model parameters, such that $\phi(r)$ is dimensionless (having therefore canonical dimensions within a two-dimensional action).

Returning to our generalisation of Derrick's theorem, it is easily verified that it is satisfied for the one-dimensional cases described in Secs.~\ref{section:quadscreening} and~\ref{section:linscreening}. For the quadratically coupled model, we have from Eq.~\eqref{eq:Derricksource1} that
\begin{equation}
    \frac{m_s}{2\mu^2M^2}\,\varphi^2(0)=-\mathcal{E}_{K}[\varphi]+\mathcal{E}_V[\varphi]+\mathcal{E}_{\rho}[\varphi]~,
\end{equation}
where
\begin{subequations}
\begin{align}
    \mathcal{E}_K[\varphi]&=\int_{-\infty}^{+\infty}{\rm d}x\;\frac{1}{2}\left(\varphi'(x)\right)^2~,\\
    \mathcal{E}_V[\varphi]&=\int_{-\infty}^{+\infty}{\rm d}x\;\left[-\frac{1}{2}\varphi^2(x)+\frac{1}{4}\varphi^4(x)+\frac{1}{4}\right]~,
\end{align}
\end{subequations}
are the dimensionless contributions from the kinetic and potential terms, where the constant ensures that the field has zero energy density in vacuum. Their values are
\begin{equation}
    \mathcal{E}_K[\varphi]=\mathcal{E}_V[\varphi]=\frac{1}{3\sqrt{2}}\left(1-\varphi(0)\right)^2\left(2+\varphi(0)\right)~,
\end{equation}
and we therefore require
\begin{equation}
    \mathcal{E}_{\rho}[\varphi]=\int_{-\infty}^{\infty}{\rm d}x\;\frac{\rho}{2\mu^2M^2}\,\varphi^2(x)=\frac{m_s}{2\mu^2M^2}\,\varphi^2(0)~,
\end{equation}
which is indeed the case. We also see explicitly that Eq.~\eqref{eq:Derricksource2} is satisfied, since $\mathcal{E}_K[\varphi]>0$.

For the linearly coupled model, since the solution is of the same form as the quadratically coupled case, it is easily shown that Derrick's theorem is again satisfied, as long as $m_s\leq \sqrt{2}\mu M$. At the branch point in Eq.~\eqref{eq:oneDvarphi}, the energy becomes complex, signalling an instability, and Derrick's theorem breaks down. However, this feature is unique to one spatial dimension.

In this work, Eqs.~\eqref{eq:Derricksource1} and~\eqref{eq:Derricksource2} will be used to perform quantitative checks of the numerically obtained field profiles (see Sec.~\ref{section:numerics}). However, as we have seen, they provide us with a qualitative prediction for the behaviour of those numerical solutions in the point-like limit in $d=3$. 


\section{Computing the field profiles around extremely compact sources}
\label{sec:fieldprofiles}

In this section, we will describe how to compute the static scalar field profiles around sources as they become more compact and point-like. We begin by describing how the piecewise analytic approach to solving the field profiles can be extended to more compact sources.  We then proceed to describe our numerical approach to computing these profiles, and the properties of the solutions that we find. We will consider solutions around an infinite cylinder ($d=2$) and around a sphere ($d=3$). Both sources will have a radius $r_s$ and uniform density $\rho$, such that $\rho_{\rm in}=\rho$ inside the source ($r\leq r_s$) and $\rho_{\rm out}=0$ outside the source ($r>r_s$). We note that $\rho$ is a mass density with units mass/length$^3$ for a sphere and units mass/length$^2$ for an infinite cylinder. 

Since we consider extended sources of finite density, we demand that the field profile is regular at $r=0$ and that it approaches one of the field values that minimise the effective potential as $r \rightarrow \infty$. In vacuum, the effective potential for both models is invariant under $\phi \rightarrow -\phi$, and the field is equally likely to sit in the minimum at positive or negative $\phi$. If the field couples to matter quadratically then the symmetry under $\phi \rightarrow -\phi$ is preserved in the matter coupling, and, up to an overall sign, the form of the field profile is insensitive to the choice of vacuum at $r\rightarrow \infty$. 

The linearly coupled model requires more careful consideration, as the symmetry is broken explicitly inside the source by the matter coupling.  For certain combinations of boundary conditions and sources, we will be unable to find piecewise analytic solutions. For example, if the boundary condition is $\phi \to -\bar{\phi}_{\infty}$ as $r\to\infty$ then the field may enter a regime close to the surface of the source where the $\phi^4$ term of the full effective potential dominates.  It will not be possible to find a piecewise analytic solution in this case. 

In what follows, and for the linearly coupled model, we will consider configurations where the field is in the $\bar{\phi}_{\infty}$ vacuum at infinity but where the field becomes large and negative inside the source.  This raises the question of whether such a configuration is truly stable or whether it can tunnel to nucleate a bubble of the $-\bar{\phi}_{\infty}$ vacuum around the source.  We leave further investigation of this interesting possibility for future work. 


\subsection{Piecewise approximations}
\label{section:piecewisepotentials}

In Sec.~\ref{section:models}, we outlined the approach that is common in the literature to finding piecewise analytic solutions for the scalar field profiles around compact objects.  For the quadratically coupled model, we divided space into two parts:~one inside and one outside the source; for the linearly coupled model, we further divided the space inside the source into two regions.  We will refer to these as two- and three-part solutions, respectively.  
When we outlined these approximations to the field profiles and the corresponding screening factors, however, we failed to check whether the values of the field in each spatial region were consistent with the approximations made to the effective potential.  It is easy to see that for more point-like sources, these approximations do fail, and the solutions in Sec.~\ref{section:models} are inconsistent. For a wide range of sources, however, we can continue to make successive linear and quadratic approximations to the scalar potential, before matching the solutions to these approximate equations together.

The effective potentials in Eqs.~\eqref{eq:quadpotential} and~\eqref{eq:linpotential} depend on  three fixed scales, which we  take to be the Compton wavelength $1/\mu$ (in vacuo), the vacuum expectation value (vev) $\bar{\phi}_{\infty} = \mu/\sqrt{\alpha}$ and the matter coupling $\rho/M^n$ ($n=1,2$, depending on whether the coupling to matter is linear or quadratic in the scalar field). This gives us the freedom to make two rescalings of the equations of motion and express them only in terms of dimensionless quantities. The solutions will depend  on the dimensionless distance $x = \mu r$ (so that the source radius is $x_s=\mu r_s$) and the dimensionless source density $\tilde{\rho} \coloneqq \rho\bar{\phi}_{\infty}^{n-2}/\mu^2 M^n$. For the quadratic coupling, this dimensionless density is $\tilde{\rho} = \rho/\mu^2 M^2$; it is the ratio of the source density to the critical density. For the linear coupling, it is $\tilde{\rho} = \sqrt{\alpha}\rho/\mu^3 M$.

When taking the point-source limit $x_s \to 0$, we keep the source mass  fixed.  We therefore parametrise the source mass as $\tilde{m}_s \coloneqq \Omega_d\tilde{\rho}x_s^d/d = m_s \mu^{d-2}\bar{\phi}_{\infty}^{n-2}/M^n$, where $\Omega_d$ is the $d$-dimensional solid angle defined in Eq.~\eqref{eq:solidangle}. For the quadratic coupling, this dimensionless mass is $\tilde{m}_s =  m_s \mu^{d-2}/M^2$; for the linear coupling, it is $\tilde{m}_s =  m_s \sqrt{\alpha}\mu^{d-3}/M$.
Solutions and screening factors then depend only on $\tilde{m}_s$ and $x_s$, except for an overall factor of $\bar{\phi}_{\infty}^2/M^2$ for the screening factors in the quadratically coupled models (Sec.~\ref{section:quadscreening}).


\subsubsection{The effective potential}
\label{sec:pieceout}

We now briefly outline when linear and quadratic approximations can be made to the effective potential for our models.  We begin by considering the effective potential outside the source, which is the same for both the quadratically and linearly coupled models. 

Requiring continuity of the effective potential and its first derivative outside the source, we find that the effective potential is well approximated by
\begin{equation}
V_{\rm out}(\phi) \equiv \mu^2\begin{cases}
 -\frac{1}{2}\phi^2, \qquad & 0 \le |\phi| \le \frac{1}{3}\bar{\phi}_{\infty} \\
 \frac{1}{18}\bar{\phi}_{\infty}^2 - \frac{1}{3}\bar{\phi}_{\infty}|\phi|,  & \frac{1}{3}\bar{\phi}_{\infty} \le |\phi| \le \frac{5}{6}\bar{\phi}_{\infty} \\
 \frac{3}{4}\bar{\phi}_{\infty}^2 + \phi^2 - 2\bar{\phi}_{\infty}|\phi|, & \frac{5}{6}\bar{\phi}_{\infty} \le |\phi| \le \bar{\phi}_{\infty} \label{eq:3pp}
\end{cases}~.
\end{equation}
As a result, we can consider up to three spatial regions outside the source.

Inside the source, $\rho>0$, and we must differentiate between the behaviour of the linearly and quadratically coupled models. 

\begin{enumerate}
\item \textbf{Quadratic coupling.}  We approximate the effective potential as
\begin{equation}
\label{eq:Vin}
V_{\rm in}(\phi) \equiv V(0) + \frac{1}{2}\nu^2\mu^2(\phi-\bar{\phi}(\rho))^2~.
\end{equation}
This expansion is valid when $\nu$, the mass of the scalar fluctuations in units of $\mu$, satisfies $\nu^2 \gg 2\bar{\phi}(\rho)(\phi-\bar{\phi}(\rho))/\bar{\phi}_{\infty}^2$. This is always true for sources with $\rho>\mu^2 M^2$, since $\bar{\phi}(\rho)=0$ and $\nu^2=\rho/(\mu^2M^2)-1 > 0$ for this case. For sources with $\rho<\mu^2 M^2$, $\bar{\phi}(\rho) = \bar{\phi}_{\infty}\sqrt{1 - \rho/(\mu^2 M^2)}$ and $\nu^2 = 2[1 - \rho/(\mu^2M^2)]$. The condition $\nu^2 \gg 2\bar{\phi}(\rho)(\phi-\bar{\phi}(\rho))/\bar{\phi}_{\infty}^2$ then simplifies to $\phi - \bar{\phi}(\rho) \ll \bar{\phi}(\rho)$, and it is possible to make a consistent quadratic approximation around this minimum everywhere inside the source as long as $\phi$ remains less than $2\bar{\phi}(\rho) = 2\bar{\phi}_{\infty}\sqrt{1 - \rho/(\mu^2 M^2)}$. Otherwise, the field profile must be found numerically, as we do below.

\item \textbf{Linear coupling. } For the linearly coupled model, there are  two cases to consider:~$\tilde{\rho} < \sqrt{12}/9$, when the effective potential inside the source has two non-degenerate minima; and $\tilde{\rho} >\sqrt{12}/9$, when the effective potential has only one minimum. 

When the effective potential has two minima, the first minimum $\phi_{+}$ satisfies $\bar{\phi}_{\infty}/\sqrt{3} < \phi_{+} < \bar{\phi}_{\infty}$ (moving from the upper to the lower limit  of this range as the density increases). The second minimum $\phi_{-}$ lies in the range $\phi_{-} < -\bar{\phi}_{\infty}$ (decreasing monotonically as $\tilde{\rho}$ increases). The only situation that we can approximate analytically is when $\phi$ remains close to $\phi_+$ inside the source, in which case we can approximate the effective potential as a quadratic around $\phi_+$, i.e.,
\begin{equation}
\label{eq:Vout}
V_{\rm out}(\phi) \equiv V(\phi_+) + \frac{1}{2}\nu^2\mu^2(\phi-\phi_+)^2~.
\end{equation}

When only one minimum is present, the field may pass through a region where $|\phi| \ll \tilde{\rho}^{1/3}\bar{\phi}_{\infty}$ before it is able to reach the minimum of the potential. This region defines a ``shell'' near the surface of the source. In this case, we find
\begin{equation}
V(\phi) = \mu^2\begin{cases}
\frac{1}{4\bar{\phi}_{\infty}^2}\bar{\phi}(\rho)^4 - \frac{1}{2}\bar{\phi}(\rho)^2 + \tilde{\rho}\bar{\phi}_{\infty}\bar{\phi}(\rho) + \frac{1}{2}\nu^2(\phi - \bar{\phi}(\rho))^2~, & \quad \phi \le \zeta\bar{\phi}_{\infty} \\
 \tilde{\rho}\bar{\phi}_{\infty}\phi - \frac{1}{2}\phi^2~, &\quad \zeta\bar{\phi}_{\infty} \le \phi \label{eq:2pp}
\end{cases}~,
\end{equation}
where
\begin{subequations}
\begin{align}
\nu^2 &= \frac{\tilde{\rho}^2 -2U}{2U - 2\tilde{\rho}\bar{\varphi}(\rho) + \bar{\varphi}^2(\rho)}~, \\
\zeta &= \frac{2 U- \tilde{\rho}\bar{\varphi}(\rho)}{\tilde{\rho} - \bar{\varphi}(\rho)}~, \\  
U &= \frac{\alpha}{\mu^4} V_{\rm eff}(\bar{\phi}(\rho))~,
\end{align}
\end{subequations}
where $\bar{\varphi}(\rho)\coloneqq \bar{\phi}(\rho)/\bar{\phi}_{\infty}$. Although not obvious from the expression, $\zeta < 0$ by construction, since $\bar{\phi}(\rho) < \zeta\bar{\phi}_{\infty} < 0$. As $\tilde{\rho} \to \infty$, $\zeta \to -\tilde{\rho}^{1/3}/2 = \bar{\phi}(\rho)/(2\bar{\phi}_{\infty})$.

The construction of Eq.~\eqref{eq:2pp} is similar to the much simpler construction of the linearly coupled solutions in Sec.~\ref{section:linscreening}, where instead only the $\tilde{\rho}$ term is kept in the shell and the interior region has a constant $\phi(r) = \bar{\phi}(\rho)$. By Eq.~\eqref{eq:3Drs}, the equivalent of $\zeta$ is then $1 - \rho r_s^2/(2M\bar{\phi}_{\infty})$. Note that, in the case of quadratic coupling, we only have one approximation to the potential and one region inside the source; for the linearly coupled model, we have at most two.  This is the same as in the calculation of the screening factors outlined in Sec.~\ref{section:models}. However, in what follows, we will be careful to check the validity of these approximations.

\end{enumerate}


\subsubsection{Piecewise scalar field profiles}
\label{sec:profiles}

\begin{figure}[t!]
\centering
\subfloat[][\label{fig:analyticprofiles2} quadratically coupled model]{\includegraphics{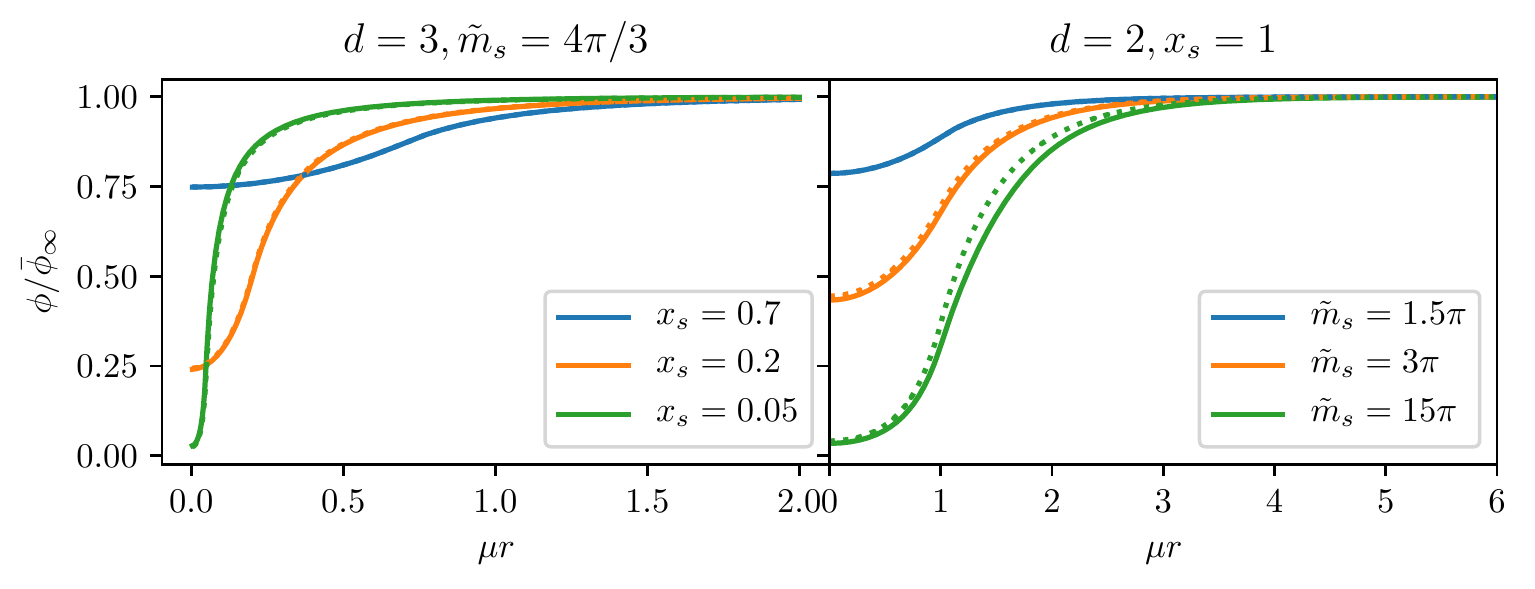}}\\
\subfloat[][\label{fig:analyticprofiles1} linearly coupled model]{\includegraphics{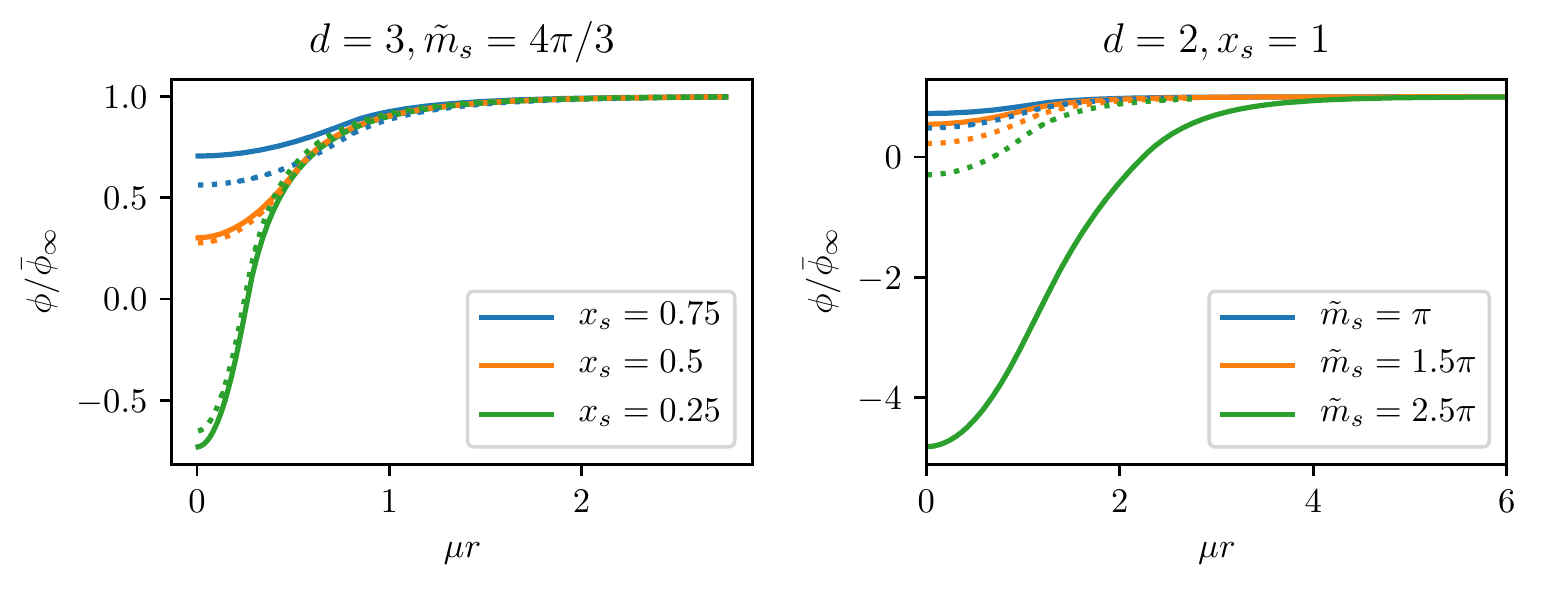}}
\caption{Examples of piecewise-constructed field profiles around spherical sources, i.e., $d=3$ (left panels), and cylindrical sources, i.e., $d=2$ (right panels), for (a) the quadratically coupled model and (b) the linearly coupled model. In each case, the values of the dimensionless source radius $x_s$ and the dimensionless source mass $\tilde{m}_s$ are chosen so that we require a two-part (solid blue), three-part (solid orange) or four-part (solid green) solution. The quartic self-coupling was taken to be unity, i.e., $\alpha=1$.  The dotted lines give the corresponding approximate solutions constructed as described in Secs.~\ref{section:quadscreening} and~\ref{section:linscreening}. In the quadratically coupled case, the solid blue and dotted blue lines are indistinguishable.  We remark that there is generally good agreement with our multi-part solutions in the quadratically coupled case. On the other hand, we see that the standard approximate solutions deviate significantly from our multi-part solutions in the linearly coupled case and especially for cylindrically symmetric sources, i.e., in $d=2$.}
\label{fig:analyticprofiles}
\end{figure}

Figure~\ref{fig:analyticprofiles} provides examples of piecewise continuous field profiles around spherically symmetric sources (in three spatial dimensions) and cylindrically symmetric sources (in two spatial dimensions) for both the quadratically (Fig.~\ref{fig:analyticprofiles2}) and linearly coupled models (Fig.~\ref{fig:analyticprofiles1}). The constants of integration in the piecewise solutions to the scalar equations of motion are determined by ensuring continuity of the field and its first derivative at the matching surfaces. An example of this construction is given in Appendix \ref{sec:example}. For cylindrically symmetric sources, analytic solutions to the equations determining the constants of integration are not available, and they must instead be solved numerically. The source parameters for the various examples shown in Fig.~\ref{fig:analyticprofiles} were chosen to give sources of sufficient compactness to require a two-part (solid blue), three-part (solid orange) or four-part (solid green) piecewise construction. The dotted lines are the ``standard'' two- (for quadratic coupling) and three-part (for linear coupling) approximations to the field profiles around a source of the same mass and radius, as outlined in Secs.~\ref{section:quadscreening} and ~\ref{section:linscreening}, and references therein. For the linearly coupled case, the sources have $\tilde{\rho} > \sqrt{12}/9$ so that the effective potential inside the source only has a single minimum. By inspection, we see that the approximations for the linear coupling fail to capture the full behaviour of the field profiles, and this worsens as the sources become more compact.

One can see from these profiles  that the qualitative behaviour is the same around spherically and cylindrically symmetric sources:~for the quadratic coupling, the field at the centre of the source tends towards zero as the source density increases; for the linear coupling, the field at the centre of the source becomes more negative as the source becomes more compact. We begin to see indications of the behaviour of the field profiles around extremely compact sources. For the quadratic coupling, the field value at the origin decreases towards  zero as the source becomes more compact.
In three spatial dimensions, as the source is made more compact, the field profile becomes flatter in the exterior of the source, and the field rises more steeply in the vicinity of the surface of the source. This matches our conclusions in Sec.~\ref{section:Derrick} that the gradient of the field should diverge at the origin, while the field itself should go to $\bar{\phi}(\rho) = 0$.  This will be confirmed by our numerical solutions in Sec.~\ref{sec:scaling}. Also in agreement with our conclusions in Sec.~\ref{section:Derrick}, we see in the linearly coupled model that the field evolves through $\phi = 0$ and becomes negative for compact sources, and our calculations indicate the possibility that the value of the field at the origin diverges in the point-source limit. We will explore this further with our numerical field profiles in Sec.~\ref{sec:scaling}. 


\subsection{Numerical calculation of field profiles}
\label{section:numerics}

The piecewise approximations to the field profiles, derived in the previous section, give us some intuition for how the quadratically and linearly coupled models behave around increasingly point-like sources. In this section, we compute  numerical solutions to the full equations of motion for both the linearly and quadratically coupled models around cylindrical and spherical sources.

To obtain them, we have modified the \phienics{} code\footnote{\url{https://github.com/scaramouche-00/phienics}}, a numerical code building on the
\fenics{} library\footnote{\url{https://fenicsproject.org/}} \cite{FEniCS_citations, New_FEniCS_book, Old_FEniCS_book} and applying the finite element method to problems of screening. The modified numerical code, as well as \texttt{Jupyter} notebooks for the main results presented in this paper, are available at \github{}. The numerical method underlying the \phienics{} code has been extensively described in Ref.~\cite{phienics_method_paper}. In the following, we limit ourselves to summarising its main qualitative aspects, while focusing on the details of the implementation specific to this work.  We conclude by showing sample field profiles around highly compact sources.


\subsubsection{Equations of motion}
\label{sec:numim}

We solve the  static, spherically or cylindrically symmetric equations of motions:
\begin{itemize}
\item Quadratic matter coupling:
\begin{equation}\label{Eq:EoM_qua}
\nabla^2\phi + \mu^2\phi - \alpha\phi^3 - \frac{\rho}{M^2}\phi = 0,
\end{equation}
\item Linear matter coupling:
\begin{equation}\label{Eq:EoM_lin}
\nabla^2\phi + \mu^2\phi - \alpha\phi^3 - \frac{\rho}{M} =0.
\end{equation}
\end{itemize}
for field profiles with the boundary conditions $\{{\rm d}\phi(r\rightarrow0)/{\rm d}r=0; \phi(r\rightarrow\infty)=\mu/\sqrt{\alpha} \}$.

For computational convenience, we  work with dimensionless quantities:~further details can be found in Appendix A of Ref.~\cite{phienics_method_paper}. However, in this section, we will express all equations in their native form for readability.
 
The equations of motion can trivially be rewritten in terms of $\psi= \phi - \bar{\phi}_{\infty}$, for which the  Dirichlet boundary condition $\phi(r\rightarrow\infty)=\mu/\sqrt{\alpha} $ becomes $\psi(r\to\infty) =0$. We have developed both solvers, i.e., one for the field $\phi$ and another one for the field $\psi$. We find the latter to be more accurate, thanks to its better characterisation of the exponential decay at large $r$.

We solve the non-linear Eqs.~\eqref{Eq:EoM_qua} and \eqref{Eq:EoM_lin} using  Newton's method, an iterative technique that aims to find successively improved approximations $\{ \phi^{(k)}, k=1,\cdots,n_{\rm iter}\}$ to the true solution $\phi$, given some sufficiently close initial guess $\phi^{(0)}$, where $n_{\rm iter}$ is the number of iterations. If $\phi^{(k)}$ is sufficiently close to the true solution $\phi$, we can expand the non-linear terms in the equation of motions to first order, so that they become linear equations in $\phi$. For the case of quadratic coupling, we can approximate  Eq.~\eqref{Eq:EoM_qua} at some iteration $k+1$ by
\begin{equation} \label{Eq:Newton_iteration}
  \nabla^2\phi^{(k+1)} + \mu^2\phi^{(k+1)} - 3\alpha (\phi^{(k)})^2 \phi^{(k+1)} - \frac{\rho}{M^2}\phi^{(k+1)}  +2\alpha (\phi^{(k)})^3=0~,
\end{equation}
and, for linear coupling, the solution to Eq.~\eqref{Eq:EoM_lin} can be found from
\begin{equation} \label{Eq:Newton_iteration2}
  \nabla^2\phi^{(k+1)} + \mu^2\phi^{(k+1)} - 3\alpha (\phi^{(k)})^2 \phi^{(k+1)} +\alpha (\phi^{(k)})^3 -\frac{\rho}{M}=0~.
\end{equation}
We solve the discretised form of Eqs.~\eqref{Eq:Newton_iteration} and \eqref{Eq:Newton_iteration2}, as described in the next subsection, to obtain an improved approximation $\phi^{(k+1)}$, which will be used for the subsequent iteration.
The $\{ \phi^{(k)} \}$ will not exactly satisfy the continuum Eqs.~\eqref{Eq:Newton_iteration} or \eqref{Eq:Newton_iteration2}, but the residuals will decrease with the iterations until a convergence criterion has been met, or the solution has saturated the accuracy that the scheme is capable of achieving. The convergence criteria that we use are discussed in Ref.~\cite{phienics_method_paper}.


\subsubsection{Spatial discretisation:~the finite element method}

Equations \eqref{Eq:EoM_qua} and \eqref{Eq:EoM_lin} are non-linear boundary value problems, which are generally challenging to solve numerically. For cases of physical interest, this difficulty is supplemented by the presence of a large hierarchy in the ratio of the field's Compton wavelength to the source radius. Moreover, as we need to consider a step source profile in order to compare against the conditions from Derrick's theorem, we can anticipate the need to resolve sharp transitions, potentially inducing numerical shocks.

The finite element method is a powerful technique that is well-suited to these challenges.  It has been applied successfully to the study of chameleon screening in Refs.~\cite{Upadhye:2006vi, chameleon_shape_dep, Sabulsky:2018jma}, symmetron screening in Refs.~\cite{Jaffe:2016fsh, Brax:2018zfb, Elder:2019yyp} and Vainshtein screening in Refs.~\cite{phienics_method_paper, Kuntz:2019plo, Vainshtein_paper}. A detailed presentation of the method is beyond the scope of this work: we refer the interested reader to Refs.~\cite{langtangen2019introduction, Old_FEniCS_book}, while listing here the most relevant basic aspects. A discussion focused on using the finite element method for solving equations of motion with screening, including worked examples, is given in Ref.~\cite{phienics_method_paper}.

The finite element method works by solving a discrete approximation to the original continuous boundary value problem. We consider a finite interval of the radial domain $r/r_s \in [0,\hat{r}_{\rm max}]$, where $\hat{r}_{\rm max}$ is taken to be $\hat{r}_{\rm max}=10^4/x_s$ (so that $\hat{r}_{\rm max} r_s = 10^4/ \mu$, i.e., $10^4$ times the field's Compton wavelength) when $x_s < 1$, and $\hat{r}_{\rm max} = 100$ when $x_s \geq 1$. This domain is discretised into a collection of $N$ intervals $\{[\hat{r}_i,\hat{r}_{i+1}], i \in 0,\cdots,N \}$, with $\hat{r}_{i=0}=0$ and $\hat{r}_{i=N+1}=\hat{r}_{\rm max}$, which we will refer to as ``cells''. We will additionally refer to the collection of all cells as a ``mesh'' and to cell extrema as mesh ``vertices''. Details of the meshes that we use are given in Ref.~\cite{phienics_method_paper}.

A key aspect of the finite element method is that cells can be of varying size. In other words, mesh vertices do not need to be equally spaced. This is particularly helpful in addressing both the hierarchy between the Compton wavelength and the source radius, and the expected sharp variation in the field in the vicinity of the source-vacuum transition. This flexibility in placing the vertices means that we can choose to have finer resolution around the source radius and a coarser grid far away from the source, where the field is expected to vary slowly. If we were restricted to a regular grid, obtaining high resolution at the source-vacuum transition whilst simultaneously covering a large box would result in a large number of vertices to solve for and correspondingly large computational costs.

On every mesh cell $i$, one then defines a discretised function space $S_i$, spanned by a number $o$ of basis functions. In this work, our basis functions will be Lagrange polynomials.  On every cell, a function can then be represented in terms of the coefficients of its expansion in the chosen basis. By imposing that the function values at mesh vertices coincide for contiguous cells, one can then choose to obtain a piecewise continuous function on the whole discretised domain\footnote{It is not always desirable to employ globally continuous function spaces; see, e.g., Ref.~\cite{phienics_method_paper}.}. Solving the original boundary value problem then amounts to determining the coefficients of the expansion on all cells. For the order $o$ used in the results we present, as well as for any other numerical setting, we refer the reader to the code and the \texttt{Jupyter} notebooks on \github{}.


\subsubsection{Weak form of the equations}

The finite element method is particularly well suited for the ``weak'' (i.e., integral) formulation of the field equations. Consider the equation $\mathcal{D}[f]=0$, where $\mathcal{D}$ is a differential operator. In the weak formulation of this equation, one does not ask that the equation of motion hold at every point in the equation domain (i.e., the ``strong'' form), but rather that the inner product $\langle\mathcal{D}[f],v\rangle$ vanish for all ``test'' functions $v$ in a chosen function space.

In this work, we are interested in spherically or cylindrically symmetric functions, and we use the inner product:
\begin{equation}
 \langle f, g \rangle = \int_0^{\infty} {\rm d}r \; r^{d-1} f g,
\end{equation}
which is the Euclidean inner product in $d$ dimensions except for a multiplicative geometrical factor. For the test space, we use the space of square-integrable functions on $\mathbb{R}^d$, whose gradients are also square-integrable on $\mathbb{R}^d$. This integral form is particularly adept at handling sharp transitions like the step source profile for which we intend to solve.

In this form, the Newton iteration for the quadratically coupled model in Eq.~\eqref{Eq:Newton_iteration} becomes
\begin{equation}
\begin{split}
-\int{\nabla\phi^{(k+1)}\; \nabla v\;  r^{d-1} \, {\rm d}r} + \int{\left( \mu^2 - 3\alpha(\phi^{(k)})^2 - \frac{\rho}{M^2}\right)\phi^{(k+1)} \; v \; r^{d-1} \, {\rm d}r}
\\ = -2 \alpha \int{(\phi^{(k)})^3 \; v \; r^{d-1} \, {\rm d}r}~,
\end{split}
\end{equation}
where we have used the Neumann boundary condition ${\rm d}\phi(r\rightarrow0)/{\rm d}r=0$ and integrated by parts the term $\int\nabla^2\phi^{(k+1)}\;v\;r^{d-1}\,{\rm d}r$ to lower the order of the first differential operator. Similarly, for the linearly coupled model, Eq.~\eqref{Eq:Newton_iteration2} becomes
\begin{equation}
\begin{split}
-\int{\nabla\phi^{(k+1)}\; \nabla v\;  r^{d-1} \, {\rm d}r} + \int{\left( \mu^2 - 3\alpha(\phi^{(k)})^2 \right)\phi^{(k+1)} \; v \; r^{d-1} \, {\rm d}r}
\\ = \int{\left(\frac{\rho}{M} -2 \alpha(\phi^{(k)})^3 \right)\; v \; r^{d-1} \, {\rm d}r}~.
\end{split}
\end{equation}


\subsubsection{Initial guess}
\label{sec:main-guess}

In Sec.~\ref{section:piecewisepotentials}, we showed how to approximate the scalar field profile around a compact source, which could  serve as an initial guess for our iterative Newton's solver. We use the conditions on the matching radii, derived for the piecewise approximations of Sec.~\ref{section:piecewisepotentials} to determine how many piecewise approximations we need to match together to find a good initial guess for a given set of model and source parameters. Whilst it would be possible to also use the analytic expressions previously determined as piecewise approximations in these regions, we find it numerically more efficient to use \fenics{} to solve the linearised equations of motion in each region in order to find a good initial guess.  The resulting initial guesses are an excellent match to our analytically derived piecewise profiles. The coefficients of the linearised equations of motion solved by \fenics{} to find the required initial guess are given in Appendix \ref{Sec:Guess}.

For the quadratically coupled model, when $\tilde{\rho}=1$, the approximation described in Sec.~\ref{section:piecewisepotentials} fails. In this case, we artificially detune the physical parameters away from the critical point $\rho=(\mu M)^2$ for the purpose of obtaining an initial guess for the algorithm. Further details are given in  Appendix \ref{Sec:Guess}. We stress that the initial guess only needs to be accurate enough to allow the solver to converge to the correct physical solution; it need not be an accurate reflection of the final profile, although a more accurate initial guess will ensure faster convergence to the true solution.


\subsubsection{Field profiles}

To ensure the validity of  our numerical solutions, we perform all the convergence tests described in Appendix B of Ref.~\cite{phienics_method_paper}. Moreover, we perform the Derrick's theorem tests of Eqs.~\eqref{eq:Derricksource1} and~\eqref{eq:Derricksource2}.
We have checked that our results do not depend on the specific source profile used for the density:~a step function, a smoothed top-hat profile or a Gaussian profile.
We present results for the step function in this article, since these can be compared with the analytic results in Eqs.~\eqref{eq:Derrick1} and~\eqref{eq:Derrick2}, derived from the generalisation of Derrick's theorem.

\begin{figure}[ht!]
\centering
\subfloat[][quadratically coupled model]{\includegraphics[scale=0.97]{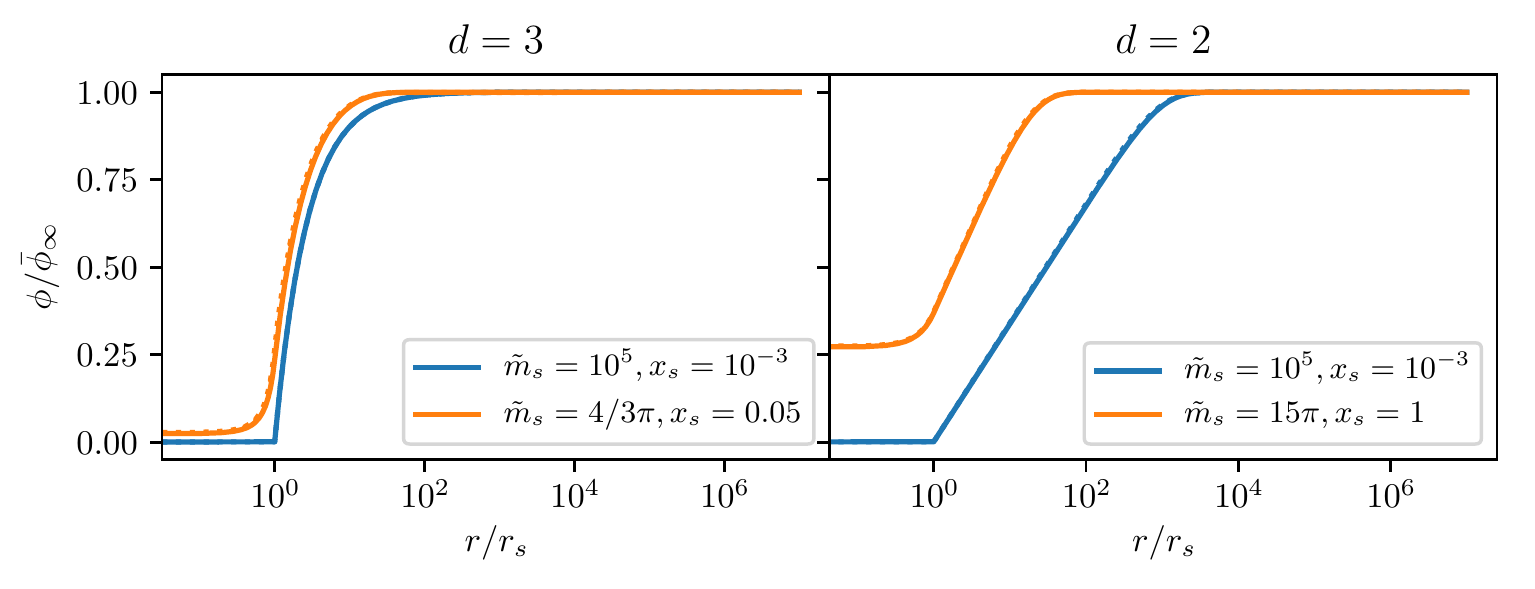}}\\
\subfloat[][linearly coupled model]{\includegraphics[scale=0.97]{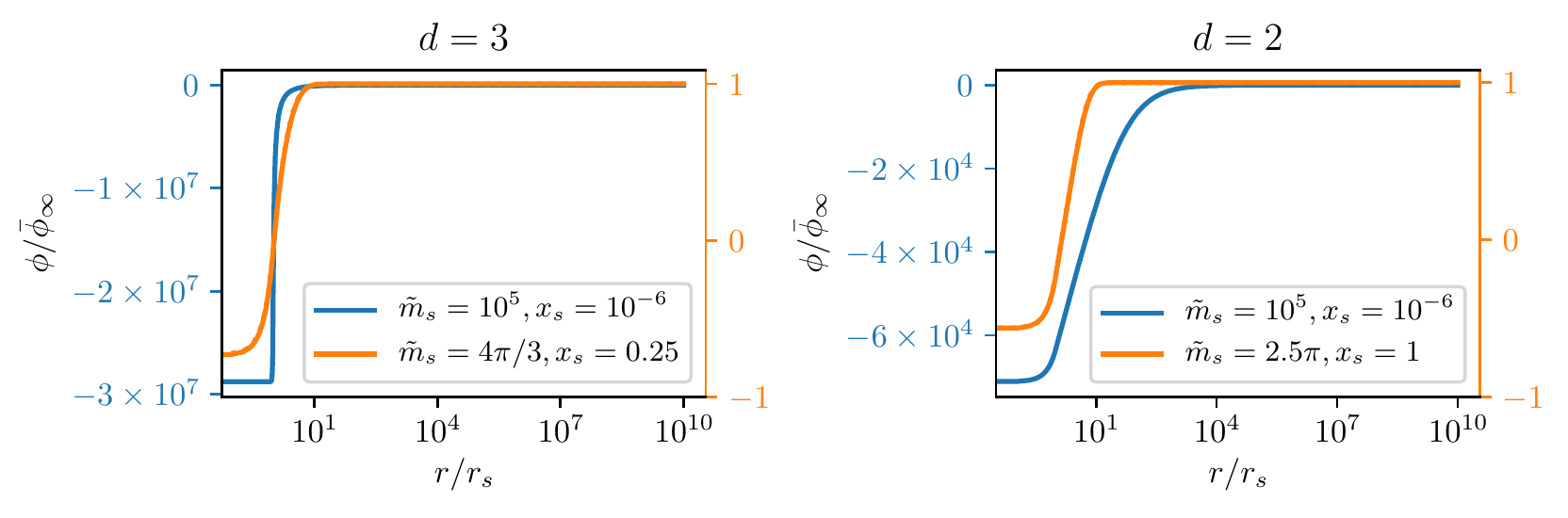}}
\caption{Comparison of the numerically obtained (solid lines) and four-part piecewise constructed field profiles (dotted lines) for (a) the quadratically coupled model and (b) the linearly coupled model for spherically ($d=3$) and cylindrically ($d=2$) symmetric sources. For the compact source in the linearly coupled case (blue), only the numerical solution is plotted, since the piecewise approximate solution overshoots the minimum to large negative values. In this case, the field value $|\phi(r_{\rm s})|$ is far greater than the vev $\bar{\phi}_{\infty}$ and is therefore outside the range of validity for the linearised potential in Eq.~\eqref{eq:3pp}. Nevertheless, the numerical results agree with our expectations, such that the field reaches a minimum value with $\phi/\bar{\phi}_{\infty}=-\tilde{\rho}^{1/3}$ at the centre of the source. This relationship holds for other parameter choices, allowing us to confirm that inside dense sources, the field generally reaches the minimum of the effective potential at the origin. Elsewhere, we see that there is excellent agreement between the full numerical profiles and the four-part piecewise constructed profiles.}
\label{fig:numericprofiles}
\end{figure}

For choices of parameters where we trust the validity of our piecewise initial guesses, such as those shown in Fig.~\ref{fig:analyticprofiles} and the orange lines in Fig.~\ref{fig:numericprofiles},  we find that the field profile found by our solver is always close to that initial guess.  For the quadratically coupled model around spherically symmetric sources (or in three spatial dimensions), we find the limiting behaviour that $\phi(r_s) \to 0$ for more compact sources, as we might expect. For cylindrically symmetric sources (or in two spatial dimensions), we find the same limiting behaviour, but shallower gradients, with the field reaching zero more slowly, cf.~Table~\ref{table:scalings}. For the linearly coupled model around spherically symmetric (three-dimensional)  sources, we find that $\phi(r=0) \approx \bar{\phi}(\rho) = -\bar{\phi}_{\infty}\tilde{\rho}^{1/3}$ for small $r_s$. For cylindrically symmetric ($d=2$) sources, we find that that $\phi(r=0)$ becomes very negative, although the particular scaling with the source parameters is more complicated, as we describe in Sec.~\ref{sec:scaling}. This  confirms our hypothesis that the field at the origin diverges near a point source in the linearly coupled model. Examples supporting all the above inferences are shown in Fig.~\ref{fig:numericprofiles}. 


\section{Implications for screening around extremely compact sources}
\label{sec:implications}

In this section, we describe the key implications of our results for the screening of dense objects whose spatial extent is significantly smaller than the Compton wavelength of the fifth-force mediator.  We summarise these implications in two ways. First, we provide a set of scaling relationships, which capture the behaviour of the scalar field and its boundary conditions at the surface of the source. These allow our results to be extrapolated to even more compact sources. Second, we compare the screening factors obtained from our full numerical solutions with those obtained from the piecewise approximations usually applied in the literature.  Where these screening factors disagree, our full numerical results provide refined screening conditions, which will find utility in phenomenological studies.  Finally, and before offering our conclusions, we draw attention to the potential relevance of quantum corrections.

\subsection{Scaling relationships}
\label{sec:scaling}

\begin{table}[t]
\centering
 \begin{tabular}{ | S{>{\centering\arraybackslash}m{0.15\textwidth}} | S{>{\centering\arraybackslash}m{0.15\textwidth}} || S{>{\centering\arraybackslash}m{0.32\textwidth}} | S{>{\centering\arraybackslash}m{0.32\textwidth}} |@{}m{0pt}@{}} 
 \hline
 Coupling & Dimensions & $\displaystyle{\left.\frac{\phi}{\bar{\phi}_{\infty}}\right|_{r\,=\,r_s}}$ & $\displaystyle{\left.\frac{\mathbf{r}\cdot\bm{\nabla}\phi}{\bar{\phi}_{\infty}}\right|_{r\,=\,r_s}}$&\\[3ex]
 \hline\hline
 \multirow{2}{*}[-0.3em]{$\displaystyle{\mathcal{L}\supset \frac{\rho}{M}\,\phi}$} & $d=3$ & $\displaystyle{-\frac{0.39\tilde{m}_s^{1/3}}{x_s}}$ & $ \displaystyle{\frac{0.17\tilde{m}_s^{2/3}}{x_s}}$\\
 \cline{2-4}
 & $d=2$ & $\displaystyle{-\frac{8.51\tilde{m}_s}{ x_s^{0.06}[1+2.05(\tilde{m}_sx_s)^{0.59}]}}$ & $\displaystyle{\frac{0.19\tilde{m}_s}{1+0.54(\tilde{m}_s x_s)^{0.39}}}$ \\
 \hline
 \multirow{2}{*}[-0.3em]{$\displaystyle{\mathcal{L}\supset \frac{\rho}{2M^2}\,\phi^2}$} & $d=3$ & $\displaystyle{1-\frac{1}{1+2.16(x_s/\tilde{m}_s)^{0.50}}}$ & $\displaystyle{1+0.67 x_s}$ \\
 \cline{2-4}
 & $d=2$ & $\displaystyle{1-\frac{1}{1+ (x_s^{0.16}/\tilde{m}_s^{0.53})}}$ & $\displaystyle{0.25\,x_s^{1-0.44}
  \times 10^{0.44\sqrt{\log^2 x_s +1.86}}}$ \\
 \hline
\end{tabular}
\caption{Summary of scaling relationships for $\phi$ and $\mathbf{r}\cdot\bm{\nabla}\phi$ at the surface of compact sources ($r=r_s$), expressed in terms of the dimensionless source radius $x_s$ and the dimensionless source mass $\tilde{m}_s$ (see Table~\ref{tab:notation}). We reiterate that ``two dimensional'' versus ``three dimensional'' refers here to cylindrically and spherically symmetric problems, respectively. For the linearly coupled model, the scaling relations hold for $x_s \in [ 10^{-6}, 1 ]$, $\tilde{m}_s \in [10^{1.5},10^6]$. For the quadratically coupled model in $d=3$, the ranges are $x_s \in [10^{-3}, 10^{-0.5}]$, $\tilde{m}_s \in [10^{1.5}, 10^3]$. For the quadratically coupled model in $d=2$ and the scaling relation for $\displaystyle{\left.\phi / \bar{\phi}_{\infty} \right|_{r\,=\,r_s}}$, the ranges are $x_s \in [ 10^{-3}, 10^{-2} ]$, $\tilde{m}_s \in [ 10^{1.5}, 10^3 ]$. For the quadratically coupled model in $d=2$ and the scaling relation for $\displaystyle{\left.\mathbf{r}\cdot\bm{\nabla}\phi / \bar{\phi}_{\infty} \right|_{r\,=\,r_s}}$, the ranges are $x_s \in [ 10^{-3}, 10^{-0.5}  ]$, $\tilde{m}_s \in [ 10^{1.5}, 10^3 ]$. Coefficients that have been determined numerically are given to two decimal places. }
\label{table:scalings}
\end{table}

The piecewise approximations to the field profiles derived in Sec.~\ref{sec:fieldprofiles}, along with the inferences from applying our generalisation of Derrick's theorem, indicated the possible behaviour of the field profiles as the compactness of the sources increases.  By complementing these results with full numerical solutions, we are able to analyse behaviour in regimes beyond the applicability of the piecewise approximations.  By these means, we can extract a set of scaling relationships that characterise how the field value $\phi(r_s)$ and  the field gradient ${\rm d}\phi(r_s)/{\rm d}r$ at the surface of the source vary with source mass and radius. This allows us to see the scaling in both the point-like limit, where $x_s\rightarrow 0$ holding $\tilde{m}_s$ fixed, and the compact limit, where $x_s \rightarrow 0$ and $\tilde{m}_s \rightarrow \infty$.    These scaling relationships for compact sources are presented in Table~\ref{table:scalings}.  Together, they provide a guide to the boundary conditions that should be applied at the surface of the source, allowing our results to be extrapolated to still more compact sources without the need to solve for the full behaviour of the field.

The key observations can be summarised as follows:
\begin{itemize}
    \item[] {\bf Linear coupling, $d=3$:} Both the field  and its gradient diverge at the surface of the source as the source becomes more compact. We note that the value of the field at the surface of the source scales with $\rho^{1/3}$, as discussed in Section \ref{section:numerics} and Fig.~\ref{fig:numericprofiles}.
    
    \item[] {\bf Linear coupling, $d=2$:} The field diverges at the surface of the source as the source becomes more compact. The behaviour of the  gradient of the field at the surface of the source depends on the way in which the limit is taken. In the point-like limit, where $\tilde{m}_s$ is held fixed and $x_s\rightarrow 0$, the gradient of the field at the surface of the source tends to a fixed value.  If $x_s$ is held fixed and the mass of the source is increased, the gradient of the field at the surface of the source diverges.
    
    \item[] {\bf Quadratic coupling, $d=3$:} The field tends to zero at the surface of the source as the source becomes more compact. In the compact limit, the gradient of the field at the surface of the source diverges as $1/r_s$, in agreement with our expectations from Derrick's theorem. 
    
    \item[] {\bf Quadratic coupling, $d=2$:} The field tends to zero at the surface of the source as the source becomes more compact. The field gradient is independent of the source mass and scales as a function of the source radius only.
    
\end{itemize}
The above behaviours agree with the inferences drawn from Derrick's theorem in Sec.~\ref{section:Derrick} in three spatial dimensions. Derrick's theorem does not allow similar inferences to be drawn in two spatial dimensions. We note that in the limit that the source becomes point-like, the field profiles may not be continuous functions, but instead be distributions, reflecting the nature of the delta-function source. We also note that, for $d>1$, we find no indication of the breakdown phenomenon identified in Ref.~\cite{Burrage:2018xta} in $d=1$ for the linearly coupled model (see also Sec.~\ref{section:linscreening}).


\subsection{Screening factors}

Having obtained full numerical solutions, we can determine screening factors without approximation.  This allows us to validate existing calculations based on approximate analytic solutions, such as those presented in Sec.~\ref{section:models}.

\begin{figure}[ht!]
\vspace{3em}
\centering
\subfloat[][\label{Fig:screening_factor1}Quadratically coupled model: The analytically derived, approximate screening factor becomes highly accurate for small source radii. However, large deviations occur when $\rho \sim \mu^2 M^2$, because the effective mass of the scalar goes to zero inside the source, and the non-linear terms in the potential dominate. This behaviour cannot be captured by a piecewise linear approximation to the potential, and no analytic prediction is available. The black ($d=3$) and white ($d=2$) dashed lines indicate where $\rho = \mu^2 M^2$.]{\includegraphics[width=\textwidth]{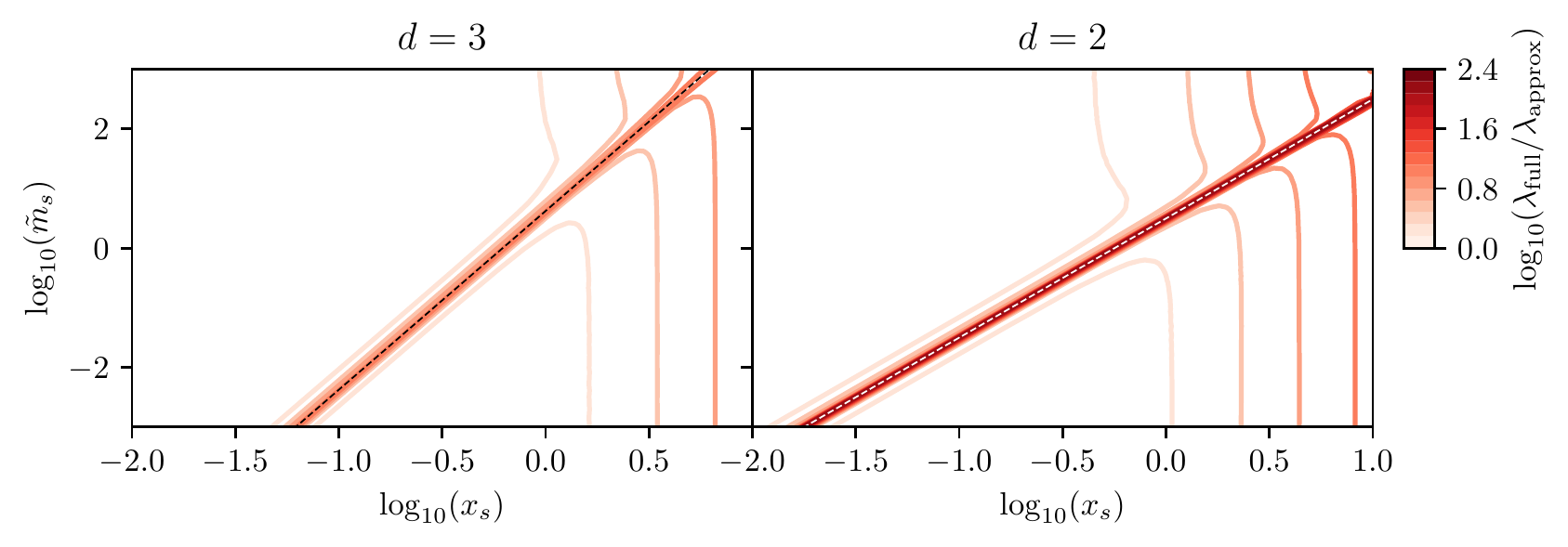}}\\
\subfloat[][\label{Fig:screening_factor2}Linearly coupled model: In both the spherically and cylindrically symmetric cases ($d=3$ and $d=2$), the full screening factor is poorly approximated by the analytically derived expression for compact sources, corresponding to the upper left regions of the plots.
]{\includegraphics[width=\textwidth]{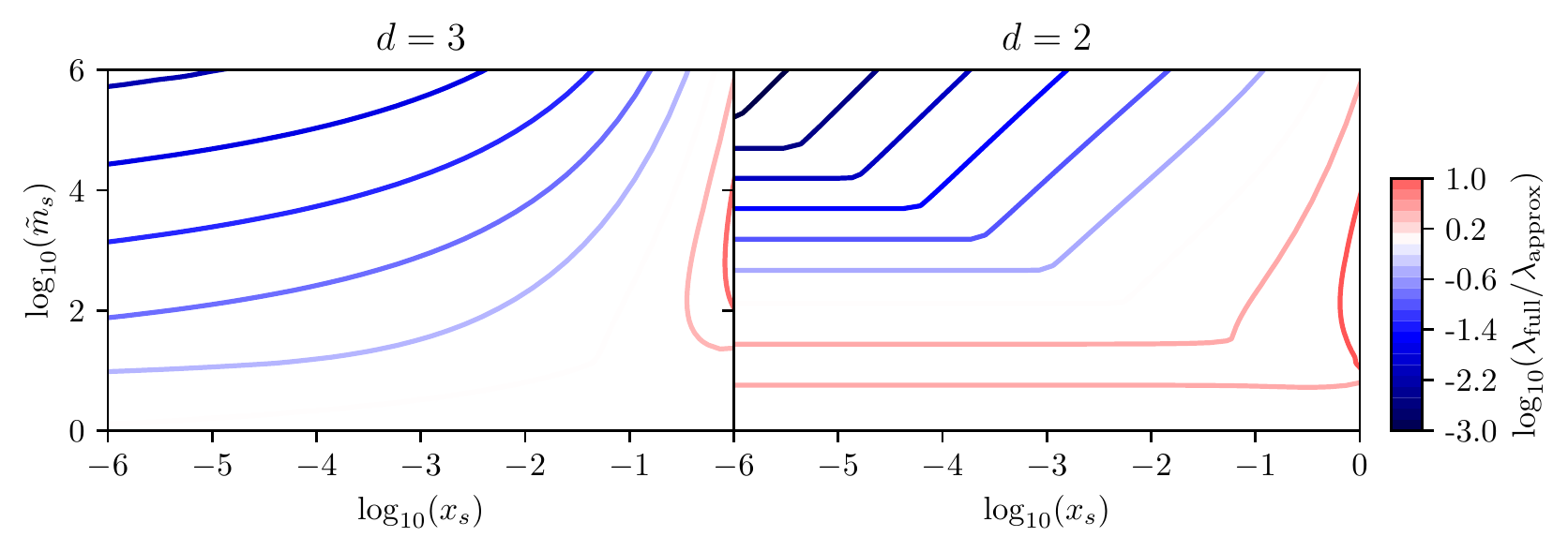}}
\caption{Logarithm of the ratio of the screening factors $\lambda_{\rm full}$ and $\lambda_{\rm approx}$, as obtained from the full and approximate field profiles for (a) the quadratically coupled model and (b) the linearly coupled model, and for spherically ($d=3$) and cylindrically ($d=2$) symmetric sources. The quartic self-coupling was taken to be $\alpha=0.1$.}\label{Fig:screening_factor}
\end{figure}

In Fig.~\ref{Fig:screening_factor}, we compare the analytic expressions for the screening factors that appear in Eqs.~\eqref{eq:3Dquadlam} and~\eqref{eq:3Dlinlam} with those obtained from Eqs.~\eqref{eq:3Dquadphiout} and~\eqref{eq:3Dlinphiout} using the field gradients from the full numerical solutions. We refer to the former as $\lambda_{\rm approx}$ and the latter as $\lambda_{\rm full}$.  We note that $\lambda_{\rm full}$ depends on the distance from the source but tends to a constant at large distances. We therefore compare the screening factors at a radius of $10/x_s$, which proves to be sufficiently distant for all cases. For positive-definite and perturbative values of the quartic self-coupling $\alpha$, we expect results qualitatively similar to those shown in Fig.~\ref{Fig:screening_factor}.

The comparisons of the screening factors can be summarised as follows:
\begin{itemize}
    \item[] {\bf Quadratic coupling (Fig.~\ref{Fig:screening_factor1}):} We find that $\lambda_{\rm full} \ge \lambda_{\rm approx}$, such that the approximate analytic solutions generally {\it overestimate} the degree of screening.  This is because the full numerical field profiles have a smaller gradient in the vicinity of the source compared to the analytic approximation.  As a result, the full numerical field profiles must have a larger gradient at greater distances from the source in order to reach the vev.  We also find that the ratio $\lambda_{\rm full}/\lambda_{\rm approx}$ depends mostly on the source radius $x_s$ in the $\rho >\mu^2M^2$ region ($\tilde{m}_s > \Omega_d x_s^d/d$), tending to 1 as $x_s \to 0$, while the source mass $\tilde{m}_s$ is only relevant for small masses. The disappearance of vertical contours for $x_s\lesssim 1$ (with the exception of the region around $\rho=\mu^2M^2$) results from the complete symmetry restoration at the centre of the source:~once the symmetry is fully restored there, a further decrease in the radius of the source has no effect.
    
    Most notable are the large deviations of $\lambda_{\rm full}$ from $\lambda_{\rm approx}$ that appear close to $\rho = \mu^2M^2$.  At this critical density, the effective mass of the scalar goes to zero inside the source.  The potential is then dominated by the $\phi^4$ term, and the analytic approximations become invalid.  This leads to a significant enhancement in the ratio $\lambda_{\rm full}/\lambda_{\rm approx}$.
    
    \item[] {\bf Linear coupling (Fig.~\ref{Fig:screening_factor2}):} While we observe contours along which $\lambda_{\rm approx} = \lambda_{\rm full}$ (in the $\tilde{m}_s > 1$ and $x_s < 1$ region of the parameter space), there is poor agreement between $\lambda_{\rm approx}$ and $\lambda_{\rm full}$ across the remainder of the parameter space.  In contrast to the quadratically coupled case and for increasingly compact sources, $\lambda_{\rm approx}>\lambda_{\rm full}$, such that the degree of screening is {\it underestimated} by the analytic approximations. In the spherically symmetric case ($d=3$), there is a transition at small $\tilde{m}_s$, where the equation for $r_{\rm shell}$ \eqref{eq:3Drs} has no real solutions and, consequently, the approximate screening factor becomes a constant $\lambda_{\rm approx} = 1$. For $x_s \to 0$, this occurs when $\tilde{m_s} < 2^{3/2}\cdot4\pi/3 \approx 11.8$.  In the cylindrically symmetric case ($d=2$), we see that the ratio of the screening factors becomes independent of the source radii for sufficiently small source mass (see Tab.~\ref{table:scalings}).

\end{itemize}

Overall, for the linearly coupled model, the analytic approximations provide a poor estimate of the degree of screening for compact sources.  On the other hand, the analytic approximations provide a reasonable reflection of the true degree of screening for the quadratically coupled model.  The important exception to this is the region of parameter space where the source nears the critical density $\rho \sim \mu^2M^2$ and for which the effective mass of the scalar inside the source goes to zero.  Interestingly, this is also the region corresponding to more diffuse sources, which may have phenomenological implications for astrophysical objects, such as nebulae.  We remark that this more strongly non-linear region of parameter space has previously been identified as interesting in the context of disk galaxies~\cite{Burrage:2016yjm, OHare:2018ayv, Burrage:2018zuj}.  However, we leave its detailed study to future work. This is not least because quantum corrections may play an important role here (see also Sec.~\ref{section:quantum}), at the very least renormalising the condition that defines the critical density.


\subsection{Quantum corrections}
\label{section:quantum}

Throughout this work, we have studied the solutions to the {\it classical} equations of motion. Before offering our conclusions, we assess the potential importance of quantum corrections. Indeed, the field gradients become large in the vicinity of point-like sources, and we might expect quantum corrections to become important. More precisely, if the classical solution is to be valid, we require the quantum fluctuations to be subdominant once smoothed over scales smaller than the characteristic length scale of the classical solution, where the latter is proportional to the Compton wavelength of the field. The arguments that follow are based on those made in Ref.~\cite{EWeinbergBook} and references therein.

Our aim is to find a smoothing length $L$ that satisfies two criteria:
\begin{itemize}
    \item[(i)] \parbox{0.9\textwidth}{\centering $L \ll \epsilon \mu^{-1}~,$}\\
    i.e., the length $L$ is smaller than the characteristic length scale of variations in the classical configuration.  Assuming that the the field rolls to the vev in about one Compton wavelength gives $\epsilon\sim 1$.
    \item[(ii)] \parbox{0.9\textwidth}{\centering $(\Delta \phi_L)_\mathrm{QM} \ll (\Delta \phi_L)_\mathrm{CL}~,$}\\
    i.e., the quantum fluctuations smoothed over an interval of length $L$ are smaller than the change in the classical solution over the same interval.
\end{itemize}
Since we will only be considering fields with non-vanishing mass in vacuum, we can always choose $L$ such that (i) is satisfied.

We note that, if the conditions (i) and (ii) are not met, it does not mean that the classical solution is invalid or uninteresting. Instead, this indicates only that one should look more closely at the predictivity of the classical approximation in this regime.

The change in the classical field configuration over the distance $L$ is approximately
\begin{equation}
    (\Delta \phi_L)_\mathrm{CL} \sim \frac{{\rm d}\phi}{{\rm d}r} L \sim \frac{\mu^2 L}{\epsilon \sqrt{\alpha}}~,
\end{equation}
where we have taken $\bar{\phi}_{\infty}/(\epsilon\mu^{-1})$ to approximate the derivative in the vicinity of the source, viz.~where the derivative is at its largest.

In order to estimate the quantum mechanical fluctuations, we first smear the quantum field over an interval of length $L$. This can be done by taking the average of the field over all space, weighted by a Gaussian of width $L$:
\begin{equation}
    \phi_L(\mathbf{x}) = \frac{1}{(2 \pi L^2)^{d/2}} \int{\rm d}^d\mathbf{y}\; e^{-(\mathbf{y} - \mathbf{x})^2/(2 L^2)} \phi(\mathbf{y})~.
\end{equation}
Placing $x$ at the origin without loss of generality, we find
\begin{align} \nonumber
    (\Delta \phi_L)_\mathrm{QM}^2 &= \bra{0} \phi_L(0)^2 \ket{0}~, \\
    &= \frac{1}{(2 \pi L^2)^d} \int{\rm d}^d\mathbf{y}\; {\rm d}^d\mathbf{z}\; e^{-(\mathbf{y}^2 + \mathbf{z}^2)/(2 L^2)} \bra{0} \phi(\mathbf{y}) \phi(\mathbf{z}) \ket{0}\nonumber\\&=\frac{1}{(2 \pi L^2)^d} \int{\rm d}^d\mathbf{y}\; {\rm d}^d\mathbf{z}\; e^{-(\mathbf{y}^2 + \mathbf{z}^2)/(2 L^2)}\int \frac{{\rm d}^d\mathbf{k}\;}{(2 \pi)^d} \frac{e^{- i \mathbf{k}\; \cdot (\mathbf{y} - \mathbf{z})}}{2 \sqrt{ \mathbf{k}^2 + \mu^2}}~.
\end{align}
After performing the spatial integrals, we are left with
\begin{align}
    (\Delta \phi_L)_\mathrm{QM}^2 &=\int \frac{{\rm d}^d\mathbf{k}\;}{(2 \pi)^d} \frac{e^{- \mathbf{k}^2 L^2}}{2 \sqrt{\mathbf{k}^2 + \mu^2}}\sim \int{\rm d}\Omega_{d}\int_0^{L^{-1}} dk \; \frac{k^{d - 1}}{\sqrt{k^2 + \mu^2}}\underset{L\mu\ll 1}{\sim} \frac{1}{L^{d-1}}~.
\end{align}

Hence, in order to satisfy the condition (ii) above, we must be able to find an $L$ such that
\begin{equation}
    \frac{\epsilon \sqrt{\alpha}} {\mu^2} \ll L^{\frac{d+1}{2}}~.
\end{equation}
In other words, we must be able to choose $L$ to be large enough that quantum fluctuations are under control. Combining this with the condition (i), we therefore require that
\begin{equation}
    \alpha \ll \frac{\epsilon^{d-1}}{\mu^{d-3}}=\begin{cases} \epsilon^2,&\qquad d=3\\ \mu\epsilon,&\qquad d=2 \end{cases}~.
\end{equation}
(We note that $\alpha$ has dimensions of mass in two spatial dimensions.)

From our analytic and numerical results, we have seen that $\epsilon \ll 1$ for small, dense sources, and our arguments from Derrick's theorem in Sec.~\ref{section:Derrick} suggest that $\epsilon \to 0$ for point sources. That is to say, although the classical solution varies rapidly in the vicinity of the point source and goes to zero, the quantum fluctuations are sufficiently large that they swamp the classical behaviour. Notice that this is not the case in $d=1$, for which we require $\alpha\ll \mu$, independent of $\epsilon$, such that we can always find a coupling small enough that the quantum effects are suppressed.  A comprehensive analysis of the quantum corrections to the scalar field profiles presented here is beyond the scope of this work.


\section{Conclusions}
\label{sec:conclusions}

The aim of this work was to study the screening of fifth forces around extremely compact sources that are effectively point-like compared to the Compton wavelength of the scalar fifth-force mediator.  To this end, we have derived a generalisation of Derrick's theorem in the presence of external sources, which allowed us to infer the behaviours of the scalar field as the compactness of these sources is increased.  These behaviours were confirmed in two and three spatial dimensions by analysing the full numerical solutions for the scalar field in models with a spontaneous symmetry breaking potential and a linear or quadratic coupling to the matter source (at the level of the action). We found that fields with a quadratic coupling are driven to zero in the vicinity of point-like sources, whereas fields with a linear coupling diverge. In addition, the numerically obtained field profiles were compared to novel piecewise analytic solutions that extend the piecewise approximations typically applied in the literature.

By these means, we have been able to establish a set of scaling relationships from which the behaviour of the scalar fields and their boundary conditions at the surface of the source can be extrapolated to other sources.  These are shown in Table~\ref{table:scalings}.  In addition, we have undertaken a critical assessment of the validity of existing screening conditions extracted from so-called screening factors.  For the linearly coupled model, this has provided a refined estimate of screening over a large region of parameter space. On the other hand, for the quadratically coupled model, we find reasonable agreement with existing screening estimates in the screened regime, except in the region of parameter space where the source has the critical density for which the field effectively become massless and the field equation becomes unavoidably non-linear.  Here, we find that existing approximations significantly overestimate the amount of screening.  Since this region of parameter space corresponds to diffuse sources, there may be potential implications for astrophysical systems, such as nebulae, which warrant further study. The comparisons between the screening factors extracted from our numerical results and those based on existing approximations are shown in Fig.~\ref{Fig:screening_factor}.

We conclude by commenting on the possible future directions highlighted by this work.  We have laid the foundations for studying the conditions under which distributions of discrete sources can be treated approximately as continuous density distributions for the purpose of analysing screening.  A natural first step towards addressing this question is the extension of this work to systems of two (or more) point-like sources, as were studied analytically in one spatial dimension in Ref.~\cite{Burrage:2018xta}.  Additionally, the results of this work should be extended to curved spacetimes and time-dependent settings. For instance, in the linearly coupled model, we identified that the explicit breaking of the $\mathbb{Z}_2$ symmetry ($\phi\to-\phi$) by the matter coupling may give rise to an instability.  As such, the increasing compactness of a collapsing matter source has the potential to induce a transition in the scalar field from one vacuum external to the source to another.  Whether this transition is realised, and the  phenomenological implications that it might have, warrants further investigation.  Finally, we have identified the regions of parameter space in which it may be necessary to account for quantum corrections to the scalar field profiles in the vicinity of compact sources, since the latter induces large spatial gradients in the field.  The treatment of these effects requires us to find the quantum corrections to spatially varying classical field profiles, and the relevant self-consistent procedures for dealing with such calculations have been developed in the context of vacuum decay~\cite{Garbrecht:2015yza, Garbrecht:2018rqx}.


\begin{acknowledgments}

The authors would like to thank Jonathan Braden, Ed Copeland, David Seery and Andrew Tolley for helpful discussions. 

We acknowledge use of the \fenics{} library \cite{Old_FEniCS_book, FEniCS_citations, New_FEniCS_book}, available from \url{https://fenicsproject.org/}.

This work was supported by a Research Leadership Award from the Leverhulme Trust [grant number RL-2016-028]; the Science and Technology Facilities Council [grant number ST/P000703/1]. PM is also supported by a Nottingham Research Fellowship from the University of Nottingham. CB is also supported by a Royal Society University Research Fellowship. DS is also supported by the European Research Council under the European Union's Horizon 2020 research and innovation programme (grant agreement No. 646702 ``CosTesGrav'').

\end{acknowledgments}

\appendix


\section{Example construction of a piecewise solution}
\label{sec:example}

In this appendix, we show the explicit construction of a piecewise analytic field profile, as discussed in Sec.~\ref{section:piecewisepotentials}.
We  consider a spherical source in the quadratically coupled model, where the conditions for smoothly matching the approximations to the scalar field profile  can  be solved analytically. We will construct here a three-part solution. We make one approximation to the form of the potential inside the source, assuming that the effective potential has the form of Eq.~\eqref{eq:Vin}, and two approximations to the potential outside the source --- the lower two approximations in Eq.~\eqref{eq:3pp}.  The field profile is then
\begin{equation}
\phi(x) = \bar{\phi}_{\infty}\begin{cases}
A\frac{\sinh(\nu x)}{x}~, &\quad x \le x_s \\
 -\frac{x^2}{18} + \frac{C_1}{x} + C_2~, & \quad x_s \le x \le x_2 \\
 1 - \frac{x_2}{6x}e^{-\sqrt{2}(x-x_2)}~, & \quad x_2 \le x
\end{cases}~,
\end{equation}
where 
\begin{subequations}
\begin{align}
A &= -\frac{2x_2^3 + 3 x_2^2(\sqrt{2}-x_s) - 18x_s + x_s^3 + x_2(3-3\sqrt{2}x_s)}{18\sinh(\nu x_s)}~,\\
C_1 &= -\frac{3x_2 + 3\sqrt{2}x_2^2 + 2x_2^3}{18}~,\\
C_2 &= 1 + \frac{\sqrt{2}x_2}{6} + \frac{x_2^2}{6}~,
\end{align}
\end{subequations}
and $x_2$ is the solution to 
\begin{equation}
 \frac{(x_s-x_2)(x_s+2x_2)-3(1+\sqrt{2}x_2)+15 x_s}{x_s(x_2-x_s)^2+3(x_2-1)(1+\sqrt{2}x_2)+15 x_s}=\frac{\tanh \nu x_s}{\nu x_s} ~.
\end{equation}
For this solution to be consistent, we must check that $\bar{\phi}_{\infty}/3 \le \phi(r_s) < 5\bar{\phi}_{\infty}/6$.


\section{The initial guess} 
\label{Sec:Guess}

In Sec.~\ref{sec:main-guess}, we described how the initial guess for the Newton's algorithm is found by numerically solving a series of linearised equations of motion. In this appendix, we give further details of that procedure. We use \fenics{} to solve the linearised equation of motion
\begin{equation} \label{Eq:num_ig}
\nabla^2 \phi - m^2\phi = b + c \rho
\end{equation}
in four spatial regions:~one inside the source and three nested shells outside the source. In these regions, the coefficients $m^2$, $b$ and $c$ in Eq.~\eqref{Eq:num_ig} are defined as follows:
\begin{center}
\begin{tabular}{|c|c|c|c|c|}
\hline
& $r < r_s$ & $r_s \leq r < r_1$ & $r_1 \leq r < r_2$ & $r_2 \leq r$ \\
\hline
$m^2$ & $\tilde{\nu}^2\mu^2$ & $- \mu^2$ & $0$ & $2 \mu^2$ \\
$b$ & $- s_b \mu^2 \bar{\phi}_{\infty}$ & $0$ & $-1/3 \, \mu^2 \bar{\phi}_{\infty}$ & $-2 \mu^2 \bar{\phi}_{\infty}$ \\
$c$ & $s_c / M$ & 0 & 0 & 0 \\
\hline
\end{tabular}
\end{center}
where the coefficients $\nu^2, s_b$ and $s_c$ are assigned the following values depending on the source density:
\begin{center}
\begin{tabular}{|c|c|c|c|}
\hline
\multicolumn{4}{|c|}{Linear coupling}\\
\hline
& $\tilde{\rho} > 1$ &  $1 \geq \tilde{\rho} > \sqrt{12}/9$  & $\tilde{\rho} \leq \sqrt{12}/9$ \\
\hline
$\tilde{\nu}^2$ & $-1$ & $2$ & $\nu^2$ \\
$s_b$ & $0$ & $2$ & $- \nu^2 \tilde{\rho}^{1/3}$ \\
$s_c$ & $1$ & $1$ & 0 \\
\hline
\multicolumn{4}{|c|}{$\tilde{\rho}= \sqrt{\alpha} \rho / ( \mu^3 M )$} \\
\hline
\end{tabular}
\quad
\begin{tabular}{|c|c|c|}
\hline
\multicolumn{3}{|c|}{Quadratic coupling}\\
\hline
& $\tilde{\rho} > 1$  &  $\tilde{\rho} < 1$ \\
\hline
$\tilde{\nu}^2$ & $\tilde{\rho} - 1$ & $2 (1 - \tilde{\rho} ) $  \\
$s_b$ &  $0$ &  $2 \left(1 - \tilde{\rho} \right)^{3/2}$ \\
$s_c$ & $0$ & $0$ \\
\hline
\multicolumn{3}{|c|}{ $\tilde{\rho} = \rho / (\mu M)^2$} \\
\hline
\end{tabular}
\end{center}
To determine the matching radii $r_1$ and $r_2$, we first  establish which of the approximating regimes described in Sec.~\ref{section:piecewisepotentials} is appropriate. After deriving the matching conditions analytically, as outlined for an example solution in Appendix \ref{sec:example}, we solve them numerically. When a two-part solution is the correct approximation, the columns  $r_s \leq r < r_1$ and $r_1 \leq r < r_2$ above are ignored. Similarly, when a three-part solution is the correct approximation, the column $r_s \leq r < r_1$ is ignored.

As we do not use the full form of Eq.~\eqref{eq:2pp}, the linearly coupled sources with $\tilde{\rho} > \sqrt{12}/9$ are split into high-density ($\tilde{\rho} > 1)$ and medium-density ($1 \geq \tilde{\rho} > \sqrt{12}/9$) regimes. For the former, we make an approximation that the matter coupling dominates over the non-linear term in the potential, such that $\nabla^2\phi = -\mu^2\phi + \rho/M$ provides us with a suitable initial guess; for the latter, we take the matter coupling to only slightly displace the field from the vev, such that $\nabla^2\phi = 2\mu^2(\phi - \bar{\phi}_{\infty}) + \rho/M$ is the equation that we use.

For the quadratically coupled model, when $\tilde{\rho}=1$, the analytic approximations described in Sec.~\ref{section:piecewisepotentials} fail:~in this case, we set $\tilde{\nu}=10^{-10}$. To obtain the matching radii, we further set $\bar{\phi}(\rho) / m_r = 10^{-10}$, where $m_r$ is a rescaling mass that we use to make the field dimensionless (see Appendix A of Ref.~\cite{phienics_method_paper}). This choice shifts the initial guess slightly away from the critical density $\rho=(\mu M)^2$, and we find it leads to fast convergence for all the parameter ranges examined in this paper.


\bibliographystyle{hieeetr}
\bibliography{refs}

\begin{thebibliography}{10}

\bibitem{Bertone:2004pz}
G.~Bertone, D.~Hooper, and J.~Silk, ``{Particle dark matter:~Evidence,
  candidates and constraints},'' {\em Phys. Rept.}, vol.~405, pp.~279--390,
  2005, arXiv:~hep-ph/0404175~[hep-ph].

\bibitem{Hui:2016ltb}
L.~Hui, J.~P. Ostriker, S.~Tremaine, and E.~Witten, ``{Ultralight scalars as
  cosmological dark matter},'' {\em Phys. Rev.}, vol.~D95, no.~4, p.~043541,
  2017, arXiv:~1610.08297~[astro-ph.CO].

\bibitem{Joyce:2014kja}
A.~Joyce, B.~Jain, J.~Khoury, and M.~Trodden, ``{Beyond the cosmological
  standard model},'' {\em Phys. Rept.}, vol.~568, pp.~1--98, 2015,
  arXiv:~1407.0059~[astro-ph.CO].

\bibitem{Akrami:2018odb}
Y.~Akrami {\em et~al.}, ``{Planck 2018 results. X. Constraints on inflation},''
  {\em Astron. Astrophys.}, vol.~641, p.~A10, 2020,
  arXiv:~1807.06211~[astro-ph.CO].

\bibitem{Englert:1964et}
F.~Englert and R.~Brout, ``{Broken symmetry and the mass of gauge vector
  mesons},'' {\em Phys. Rev. Lett.}, vol.~13, pp.~321--323, 1964.

\bibitem{Higgs:1964pj}
P.~W. Higgs, ``{Broken symmetries and the masses of gauge bosons},'' {\em Phys.
  Rev. Lett.}, vol.~13, pp.~508--509, 1964.

\bibitem{Guralnik:1964eu}
G.~S. Guralnik, C.~R. Hagen, and T.~W.~B. Kibble, ``{Global conservation laws
  and massless particles},'' {\em Phys. Rev. Lett.}, vol.~13, pp.~585--587,
  1964.

\bibitem{Shaposhnikov:2008xb}
M.~Shaposhnikov and D.~Zenhausern, ``{Scale invariance, unimodular gravity and
  dark energy},'' {\em Phys. Lett.}, vol.~B671, pp.~187--192, 2009,
  arXiv:~0809.3395~[hep-th].

\bibitem{Brax:2014baa}
P.~Brax and A.~C. Davis, ``{Conformal inflation coupled to matter},'' {\em
  JCAP}, vol.~1405, p.~019, 2014, arXiv:~1401.7281~[astro-ph.CO].

\bibitem{Ferreira:2016kxi}
P.~G. Ferreira, C.~T. Hill, and G.~G. Ross, ``{No fifth force in a scale
  invariant universe},'' {\em Phys. Rev.}, vol.~D95, no.~6, p.~064038, 2017,
  arXiv:~1612.03157~[gr-qc].

\bibitem{Burrage:2018dvt}
C.~Burrage, E.~J. Copeland, P.~Millington, and M.~Spannowsky, ``{Fifth forces,
  Higgs portals and broken scale invariance},'' {\em JCAP}, vol.~1811, no.~11,
  p.~036, 2018, arXiv:~1804.07180~[hep-th].

\bibitem{Adelberger:2009zz}
E.~G. Adelberger, J.~H. Gundlach, B.~R. Heckel, S.~Hoedl, and S.~Schlamminger,
  ``{Torsion balance experiments:~a low-energy frontier of particle physics},''
  {\em Prog. Part. Nucl. Phys.}, vol.~62, p.~102, 2009.

\bibitem{Vainshtein:1972sx}
A.~I. Vainshtein, ``{To the problem of nonvanishing gravitation mass},'' {\em
  Phys. Lett. B}, vol.~39, pp.~393--394, 1972.

\bibitem{Khoury:2003aq}
J.~Khoury and A.~Weltman, ``{Chameleon fields:~Awaiting surprises for tests of
  gravity in space},'' {\em Phys. Rev. Lett.}, vol.~93, p.~171104, 2004,
  arXiv:~astro-ph/0309300~[astro-ph].

\bibitem{Khoury:2003rn}
J.~Khoury and A.~Weltman, ``{Chameleon cosmology},'' {\em Phys. Rev.},
  vol.~D69, p.~044026, 2004, arXiv:~astro-ph/0309411~[astro-ph].

\bibitem{Mota:2006fz}
D.~F. Mota and D.~J. Shaw, ``{Evading equivalence principle violations,
  cosmological and other experimental constraints in scalar field theories with
  a strong coupling to matter},'' {\em Phys. Rev.}, vol.~D75, p.~063501, 2007,
  arXiv:~hep-ph/0608078~[hep-ph].

\bibitem{Hinterbichler:2010es}
K.~Hinterbichler and J.~Khoury, ``{Screening long-range forces through local
  symmetry restoration},'' {\em Phys. Rev. Lett.}, vol.~104, p.~231301, 2010,
  arXiv:~1001.4525~[hep-th].

\bibitem{Hinterbichler:2011ca}
K.~Hinterbichler, J.~Khoury, A.~Levy, and A.~Matas, ``{Symmetron cosmology},''
  {\em Phys. Rev.}, vol.~D84, p.~103521, 2011, arXiv:~1107.2112~[astro-ph.CO].

\bibitem{Dehnen:1992rr}
H.~Dehnen, H.~Frommert, and F.~Ghaboussi, ``{Higgs field and a new
  scalar-tensor theory of gravity},'' {\em Int. J. Theor. Phys.}, vol.~31,
  pp.~109--114, 1992.

\bibitem{Gessner:1992flm}
E.~Gessner, ``{A new scalar tensor theory for gravity and the flat rotation
  curves of spiral galaxies},'' {\em Astrophys. Space Sci.}, vol.~196, no.~1,
  pp.~29--43, 1992.

\bibitem{Damour:1994zq}
T.~Damour and A.~M. Polyakov, ``{The string dilaton and a least coupling
  principle},'' {\em Nucl. Phys.}, vol.~B423, pp.~532--558, 1994,
  arXiv:~hep-th/9401069~[hep-th].

\bibitem{Pietroni:2005pv}
M.~Pietroni, ``{Dark energy condensation},'' {\em Phys. Rev.}, vol.~D72,
  p.~043535, 2005, arXiv:~astro-ph/0505615~[astro-ph].

\bibitem{Olive:2007aj}
K.~A. Olive and M.~Pospelov, ``{Environmental dependence of masses and coupling
  constants},'' {\em Phys. Rev.}, vol.~D77, p.~043524, 2008,
  arXiv:~0709.3825~[hep-ph].

\bibitem{Hamilton:2015zga}
P.~Hamilton, M.~Jaffe, P.~Haslinger, Q.~Simmons, H.~M\"uller, and J.~Khoury,
  ``{Atom-interferometry constraints on dark energy},'' {\em Science},
  vol.~349, pp.~849--851, 2015, arXiv:~1502.03888~[physics.atom-ph].

\bibitem{Elder:2016yxm}
B.~Elder, J.~Khoury, P.~Haslinger, M.~Jaffe, H.~M\"uller, and P.~Hamilton,
  ``{Chameleon dark energy and atom interferometry},'' {\em Phys. Rev. D},
  vol.~94, no.~4, p.~044051, 2016, arXiv:~1603.06587~[astro-ph.CO].

\bibitem{Sabulsky:2018jma}
D.~O. Sabulsky, I.~Dutta, E.~Hinds, B.~Elder, C.~Burrage, and E.~J. Copeland,
  ``{Experiment to detect dark energy forces using atom interferometry},'' {\em
  Phys. Rev. Lett.}, vol.~123, no.~6, p.~061102, 2019,
  arXiv:~1812.08244~[physics.atom-ph].

\bibitem{Lemmel:2015kwa}
H.~Lemmel, P.~Brax, A.~N. Ivanov, T.~Jenke, G.~Pignol, M.~Pitschmann,
  T.~Potocar, M.~Wellenzohn, M.~Zawisky, and H.~Abele, ``{Neutron
  interferometry constrains dark energy chameleon fields},'' {\em Phys. Lett.
  B}, vol.~743, pp.~310--314, 2015, arXiv:~1502.06023~[hep-ph].

\bibitem{Cronenberg:2018qxf}
G.~Cronenberg, P.~Brax, H.~Filter, P.~Geltenbort, T.~Jenke, G.~Pignol,
  M.~Pitschmann, M.~Thalhammer, and H.~Abele, ``{Acoustic Rabi oscillations
  between gravitational quantum states and impact on symmetron dark energy},''
  {\em Nature Phys.}, vol.~14, no.~10, pp.~1022--1026, 2018,
  arXiv:~1902.08775~[hep-ph].

\bibitem{Rider:2016xaq}
A.~D. Rider, D.~C. Moore, C.~P. Blakemore, M.~Louis, M.~Lu, and G.~Gratta,
  ``{Search for screened interactions associated with dark energy below the 100
  $\mathrm{\mu m}$ length scale},'' {\em Phys. Rev. Lett.}, vol.~117, no.~10,
  p.~101101, 2016, arXiv:~1604.04908~[hep-ex].

\bibitem{Burrage:2017qrf}
C.~Burrage and J.~Sakstein, ``{Tests of chameleon gravity},'' {\em Living Rev.
  Rel.}, vol.~21, no.~1, p.~1, 2018, arXiv:~1709.09071~[astro-ph.CO].

\bibitem{Llinares:2013jza}
C.~Llinares, D.~F. Mota, and H.~A. Winther, ``{ISIS: a new N-body cosmological
  code with scalar fields based on RAMSES. Code presentation and application to
  the shapes of clusters},'' {\em Astron. Astrophys.}, vol.~562, p.~A78, 2014,
  arXiv:~1307.6748~[astro-ph.CO].

\bibitem{Winther:2014cia}
H.~A. Winther and P.~G. Ferreira, ``{Fast route to nonlinear clustering
  statistics in modified gravity theories},'' {\em Phys. Rev. D}, vol.~91,
  no.~12, p.~123507, 2015, arXiv:~1403.6492~[astro-ph.CO].

\bibitem{Winther:2015wla}
H.~A. Winther {\em et~al.}, ``{Modified gravity N-body code comparison
  project},'' {\em Mon. Not. Roy. Astron. Soc.}, vol.~454, no.~4,
  pp.~4208--4234, 2015, arXiv:~1506.06384~[astro-ph.CO].

\bibitem{Bose:2016wms}
S.~Bose, B.~Li, A.~Barreira, J.-h. He, W.~A. Hellwing, K.~Koyama, C.~Llinares,
  and G.-B. Zhao, ``{Speeding up $N$-body simulations of modified gravity:
  Chameleon screening models},'' {\em JCAP}, vol.~02, p.~050, 2017,
  arXiv:~1611.09375~[astro-ph.CO].

\bibitem{Llinares:2018maz}
C.~Llinares, ``{Simulation techniques for modified gravity},'' {\em Int. J.
  Mod. Phys. D}, vol.~27, no.~15, p.~1848003, 2018.

\bibitem{Arnold:2019zup}
C.~Arnold and B.~Li, ``{Simulating galaxy formation in $f(R)$ modified gravity:
  Matter, halo, and galaxy-statistics},'' {\em Mon. Not. Roy. Astron. Soc.},
  vol.~490, no.~2, pp.~2507--2520, 2019, arXiv:~1907.02980~[astro-ph.CO].

\bibitem{Davis:2014tea}
A.-C. Davis, R.~Gregory, R.~Jha, and J.~Muir, ``{Astrophysical black holes in
  screened modified gravity},'' {\em JCAP}, vol.~1408, p.~033, 2014,
  arXiv:~1402.4737~[astro-ph.CO].

\bibitem{Davis:2016avf}
A.-C. Davis, R.~Gregory, and R.~Jha, ``{Black hole accretion discs and screened
  scalar hair},'' {\em JCAP}, vol.~1610, no.~10, p.~024, 2016,
  arXiv:~1607.08607~[gr-qc].

\bibitem{Wong:2019yoc}
L.~K. Wong, A.-C. Davis, and R.~Gregory, ``{Effective field theory for black
  holes with induced scalar charges},'' {\em Phys. Rev.}, vol.~D100, no.~2,
  p.~024010, 2019, arXiv:~1903.07080~[hep-th].

\bibitem{Lagos:2020mzy}
M.~Lagos and H.~Zhu, ``{Gravitational couplings in chameleon models},'' {\em
  JCAP}, vol.~06, p.~061, 2020, arXiv:~2003.01038~[gr-qc].

\bibitem{Upadhye:2012qu}
A.~Upadhye, ``{Dark energy fifth forces in torsion pendulum experiments},''
  {\em Phys. Rev. D}, vol.~86, p.~102003, 2012, arXiv:~1209.0211~[hep-ph].

\bibitem{Upadhye:2012rc}
A.~Upadhye, ``{Symmetron dark energy in laboratory experiments},'' {\em Phys.
  Rev. Lett.}, vol.~110, no.~3, p.~031301, 2013, arXiv:~1210.7804~[hep-ph].

\bibitem{Pernot-Borras:2019gqs}
M.~Pernot-Borr\`as, J.~Berg\'e, P.~Brax, and J.-P. Uzan, ``{General study of
  chameleon fifth force in gravity space experiments},'' {\em Phys. Rev. D},
  vol.~100, no.~8, p.~084006, 2019, arXiv:~1907.10546~[gr-qc].

\bibitem{Stadnik:2020bfk}
Y.~V. {Stadnik}, ``{New bounds on macroscopic scalar-field topological defects
  from nontransient signatures due to environmental dependence and spatial
  variations of the fundamental constants},'' {\em Phys. Rev. D}, vol.~102,
  p.~115016, Dec. 2020, arXiv:~2006.00185~[hep-ph].

\bibitem{Brax:2013cfa}
P.~Brax, G.~Pignol, and D.~Roulier, ``{Probing strongly coupled chameleons with
  slow neutrons},'' {\em Phys. Rev.}, vol.~D88, p.~083004, 2013,
  arXiv:~1306.6536~[quant-ph].

\bibitem{Brax:2017hna}
P.~Brax and M.~Pitschmann, ``{Exact solutions to nonlinear symmetron theory:
  One- and two-mirror systems},'' {\em Phys. Rev.}, vol.~D97, no.~6, p.~064015,
  2018, arXiv:~1712.09852~[gr-qc].

\bibitem{Llinares:2018mzl}
C.~Llinares and P.~Brax, ``{Detecting coupled domain walls in laboratory
  experiments},'' {\em Phys. Rev. Lett.}, vol.~122, no.~9, p.~091102, 2019,
  arXiv:~1807.06870~[astro-ph.CO].

\bibitem{Pitschmann:2020ejb}
M.~{Pitschmann}, ``{Exact solutions to nonlinear symmetron theory: One- and
  two-mirror systems. II.},'' {\em Phys. Rev. D}, vol.~103, p.~084013, Apr.
  2021, arXiv:~2012.12752~[gr-qc].

\bibitem{Pernot-Borras:2020jev}
M.~Pernot-Borr\`as, J.~Berg\'e, P.~Brax, and J.-P. Uzan, ``{Fifth force induced
  by a chameleon field on nested cylinders},'' {\em Phys. Rev. D}, vol.~101,
  no.~12, p.~124056, 2020, arXiv:~2004.08403~[gr-qc].

\bibitem{Burrage:2018xta}
C.~Burrage, B.~Elder, and P.~Millington, ``{Particle level screening of scalar
  forces in 1+1 dimensions},'' {\em Phys. Rev.}, vol.~D99, no.~2, p.~024045,
  2019, arXiv:~1810.01890~[hep-th].

\bibitem{Vilenkin:2000jqa}
A.~Vilenkin and E.~P.~S. Shellard, {\em {Cosmic Strings and Other Topological
  Defects}}.
\newblock Cambridge University Press, 2000.

\bibitem{Derrick:1964ww}
G.~H. Derrick, ``{Comments on nonlinear wave equations as models for elementary
  particles},'' {\em J. Math. Phys.}, vol.~5, pp.~1252--1254, 1964.

\bibitem{Goldberger:2001tn}
W.~D. Goldberger and M.~B. Wise, ``{Renormalization group flows for brane
  couplings},'' {\em Phys. Rev.}, vol.~D65, p.~025011, 2002,
  arXiv:~hep-th/0104170~[hep-th].

\bibitem{deRham:2007mcp}
C.~de~Rham, ``{The effective field theory of codimension-two branes},'' {\em
  JHEP}, vol.~01, p.~060, 2008, arXiv:~0707.0884~[hep-th].

\bibitem{Burrage:2016xzz}
C.~Burrage, E.~J. Copeland, and P.~Millington, ``{Radiative screening of fifth
  forces},'' {\em Phys. Rev. Lett.}, vol.~117, no.~21, p.~211102, 2016,
  arXiv:~1604.06051~[gr-qc].

\bibitem{Coleman:1973jx}
S.~R. Coleman and E.~J. Weinberg, ``{Radiative corrections as the origin of
  spontaneous symmetry breaking},'' {\em Phys. Rev.}, vol.~D7, pp.~1888--1910,
  1973.

\bibitem{Burrage:2016rkv}
C.~Burrage, A.~Kuribayashi-Coleman, J.~Stevenson, and B.~Thrussell,
  ``{Constraining symmetron fields with atom interferometry},'' {\em JCAP},
  vol.~1612, p.~041, 2016, arXiv:~1609.09275~[astro-ph.CO].

\bibitem{Burrage:2014oza}
C.~Burrage, E.~J. Copeland, and E.~A. Hinds, ``{Probing dark energy with atom
  interferometry},'' {\em JCAP}, vol.~1503, no.~03, p.~042, 2015,
  arXiv:~1408.1409~[astro-ph.CO].

\bibitem{FEniCS_citations}
M.~S. Aln{\ae}s, J.~Blechta, J.~Hake, A.~Johansson, B.~Kehlet, A.~Logg,
  C.~Richardson, J.~Ring, M.~E. Rognes, and G.~N. Wells, ``The fenics project
  version 1.5,'' {\em Archive of Numerical Software}, vol.~3, no.~100, 2015.

\bibitem{New_FEniCS_book}
H.~P. Langtangen and A.~Logg, {\em Solving PDEs in Python}.
\newblock Springer, 2017.

\bibitem{Old_FEniCS_book}
A.~Logg, K.-A. Mardal, G.~N. Wells, {\em et~al.}, {\em Automated Solution of
  Differential Equations by the Finite Element Method}.
\newblock Springer, 2012.

\bibitem{phienics_method_paper}
J.~{Braden}, C.~{Burrage}, B.~{Elder}, and D.~{Saadeh},
  ``{{\ensuremath{\varphi}}enics:~Vainshtein screening with the finite element
  method},'' {\em JCAP}, vol.~2021, p.~010, Mar. 2021,
  arXiv:~2011.07037~[gr-qc].

\bibitem{Upadhye:2006vi}
A.~Upadhye, S.~S. Gubser, and J.~Khoury, ``{Unveiling chameleons in tests of
  gravitational inverse-square law},'' {\em Phys. Rev. D}, vol.~74, p.~104024,
  2006, arXiv:~hep-ph/0608186~[hep-ph].

\bibitem{chameleon_shape_dep}
C.~Burrage, E.~J. Copeland, A.~Moss, and J.~A. Stevenson, ``{The shape
  dependence of chameleon screening},'' {\em JCAP}, vol.~1801, no.~01, p.~056,
  2018, arXiv:~1711.02065~[astro-ph.CO].

\bibitem{Jaffe:2016fsh}
M.~Jaffe, P.~Haslinger, V.~Xu, P.~Hamilton, A.~Upadhye, B.~Elder, J.~Khoury,
  and H.~M\"uller, ``{Testing sub-gravitational forces on atoms from a
  miniature, in-vacuum source mass},'' {\em Nature Phys.}, vol.~13, p.~938,
  2017, arXiv:~1612.05171~[physics.atom-ph].

\bibitem{Brax:2018zfb}
P.~Brax, A.-C. Davis, B.~Elder, and L.~K. Wong, ``{Constraining screened fifth
  forces with the electron magnetic moment},'' {\em Phys. Rev. D}, vol.~97,
  no.~8, p.~084050, 2018, arXiv:~1802.05545~[hep-ph].

\bibitem{Elder:2019yyp}
B.~Elder, V.~Vardanyan, Y.~Akrami, P.~Brax, A.-C. Davis, and R.~S. Decca,
  ``{Classical symmetron force in Casimir experiments},'' {\em Phys. Rev. D},
  vol.~101, no.~6, p.~064065, 2020, arXiv:~1912.10015~[gr-qc].

\bibitem{Kuntz:2019plo}
A.~Kuntz, ``{Two-body potential of Vainshtein screened theories},'' {\em Phys.
  Rev. D}, vol.~100, no.~2, p.~024024, 2019, arXiv:~1905.07340~[gr-qc].

\bibitem{Vainshtein_paper}
C.~{Burrage}, B.~{Coltman}, A.~{Padilla}, D.~{Saadeh}, and T.~{Wilson},
  ``{Massive Galileons and Vainshtein screening},'' {\em JCAP}, vol.~2021,
  p.~050, Feb. 2021, arXiv:~2008.01456~[hep-th].

\bibitem{langtangen2019introduction}
H.~Langtangen and K.~Mardal, {\em Introduction to Numerical Methods for
  Variational Problems}.
\newblock Texts in Computational Science and Engineering, Springer
  International Publishing, 2019.

\bibitem{Burrage:2016yjm}
C.~Burrage, E.~J. Copeland, and P.~Millington, ``{Radial acceleration relation
  from symmetron fifth forces},'' {\em Phys. Rev. D}, vol.~95, no.~6,
  p.~064050, 2017, arXiv:~1610.07529~[astro-ph.CO].
\newblock [Erratum: Phys.Rev.D 95, 129902 (2017)].

\bibitem{OHare:2018ayv}
C.~A. O'Hare and C.~Burrage, ``{Stellar kinematics from the symmetron fifth
  force in the Milky Way disk},'' {\em Phys. Rev. D}, vol.~98, no.~6,
  p.~064019, 2018, arXiv:~1805.05226~[astro-ph.CO].

\bibitem{Burrage:2018zuj}
C.~Burrage, E.~J. Copeland, C.~K\"ading, and P.~Millington, ``{Symmetron scalar
  fields: Modified gravity, dark matter, or both?},'' {\em Phys. Rev. D},
  vol.~99, no.~4, p.~043539, 2019, arXiv:~1811.12301~[astro-ph.CO].

\bibitem{EWeinbergBook}
E.~J. Weinberg, {\em {Classical Solutions in Quantum Field Theory:~Solitons and
  Instantons in High Energy Physics}}.
\newblock Cambridge:~Cambridge University Press, 2012.

\bibitem{Garbrecht:2015yza}
B.~Garbrecht and P.~Millington, ``{Self-consistent solitons for vacuum decay in
  radiatively generated potentials},'' {\em Phys. Rev. D}, vol.~92, p.~125022,
  2015, arXiv:~1509.08480~[hep-ph].

\bibitem{Garbrecht:2018rqx}
B.~Garbrecht and P.~Millington, ``{Fluctuations about the Fubini-Lipatov
  instanton for false vacuum decay in classically scale invariant models},''
  {\em Phys. Rev. D}, vol.~98, no.~1, p.~016001, 2018,
  arXiv:~1804.04944~[hep-th].

\end{thebibliography}


\end{document}